\documentclass[%
 reprint,
nofootinbib,
 amsmath,amssymb,
 aps,
 pra,
 longbibliography,
]{revtex4-2}

\usepackage{graphicx}% Include figure files
\usepackage{dcolumn}% Align table columns on decimal point
\usepackage{amsmath,mathtools}
\usepackage{bm}% bold math
\usepackage{braket}
\newcommand\p{\partial} % Because im lazy
\usepackage{tensor}
\usepackage{hyperref}
\hypersetup{
colorlinks=true,
linkcolor=blue,
citecolor=green
}
\usepackage{siunitx} 
\DeclareSIUnit{\rps}{rps}
\AtBeginEnvironment{pmatrix}{\everymath{\displaystyle}}

\newcommand{\frameL}{\mathcal{L}}
\newcommand{\frameM}{\mathcal{M}}
\newcommand{\frameU}{\mathcal{U}}

\newcommand{\vu}[1]{\hat{\mathbf{#1}}}
\newcommand{\vb}[1]{\mathbf{#1}}

\newcommand{\gxM}{g_1'}
\newcommand{\gyM}{g_2'}

\newcommand{\gyU}{g_2''}

\newcommand{\QprM}[2]{Q_{\text{pr},#1}^{#2}}
\newcommand{\TprM}[2]{T_{\text{pr},#1}^{#2}}

\newcommand{\refl}[2]{r^{#1\prime}_{#2}}

\newcommand{\vphi}{\ensuremath{\frac{g_2'\dot{\phi}'}{c}}}

\newcommand{\kyy}{k^y_y}
\newcommand{\kyp}{k^y_\phi}
\newcommand{\kpy}{k^\phi_y}
\newcommand{\kpp}{k^\phi_\phi}
\newcommand{\myy}{\mu^y_y}
\newcommand{\myp}{\mu^y_\phi}
\newcommand{\mpy}{\mu^\phi_y}
\newcommand{\mpp}{\mu^\phi_\phi}

\newcommand{\winf}{w_\infty}
\newcommand{\ws}[1]{w_{#1}^\text{slow}}
\newcommand{\wf}[1]{w_{#1}^\text{fast}}

\newcommand{\lzero}{\lambda_0}
\newcommand{\lmax}{\lambda_\text{max}'}

\DeclareMathOperator{\erf}{erf}
\begin{document}

\preprint{APS/123-QED}

\title{Asymptotic stability of laser-driven lightsails: Enhancement by optical dispersion engineering in gratings}% Force line breaks with 

\author{Jadon Y. Lin}
\thanks{These authors contributed equally to this work.}
\affiliation{%
School of Physics, The University of Sydney, Sydney, New South Wales 2006, Australia
}%
\affiliation{%
Institute of Photonics and Optical Science, The University of Sydney, Sydney, New South Wales 2006, Australia
}%

\author{Liam van Ravenstein}
\thanks{These authors contributed equally to this work.}
\affiliation{%
School of Physics, The University of Sydney, Sydney, New South Wales 2006, Australia
}%
\affiliation{%
Institute of Photonics and Optical Science, The University of Sydney, Sydney, New South Wales 2006, Australia
}%

\author{C. Martijn de Sterke}%
\affiliation{%
School of Physics, The University of Sydney, Sydney, New South Wales 2006, Australia
}%
\affiliation{%
Institute of Photonics and Optical Science, The University of Sydney, Sydney, New South Wales 2006, Australia
}%

\author{Michael S. Wheatland}
\affiliation{%
School of Physics, The University of Sydney, Sydney, New South Wales 2006, Australia
}%

\author{Alex Y. Song}
\affiliation{School of Electrical and Computer Engineering, The University of Sydney, Sydney, New South Wales 2006, Australia}
\affiliation{%
Institute of Photonics and Optical Science, The University of Sydney, Sydney, New South Wales 2006, Australia
}%
\affiliation{The University of Sydney Nano Institute, The University of Sydney, Sydney, New South Wales 2006, Australia}

\author{Boris T. Kuhlmey}
\email{boris.kuhlmey@sydney.edu.au}
\affiliation{%
School of Physics, The University of Sydney, Sydney, New South Wales 2006, Australia
}%
\affiliation{%
Institute of Photonics and Optical Science, The University of Sydney, Sydney, New South Wales 2006, Australia
}%
\affiliation{The University of Sydney Nano Institute, The University of Sydney, Sydney, New South Wales 2006, Australia}

\date{\today}% It is always \today, today,
             %  but any date may be explicitly specified

\begin{abstract}
Lightsails are promising spacecraft that can traverse interstellar distances within decades via radiation-pressure propulsion from high-power lasers. The envisioned missions crucially rely on the sail being confined within the propelling laser beam, requiring restoring and damping mechanisms for both translational and rotational degrees of freedom. Here, we use a two-dimensional rigid model to show that full asymptotic stability of planar nanophotonic sails can be achieved through purely optical, relativistic forces and torques, which damp all unstable degrees of freedom. By judiciously optimizing the angular and frequency dispersion of diffraction gratings, we find that damping can be substantially enhanced compared to plane-mirror sails. Over the full operating band for $0.2c$ missions it is several times larger, while over narrow wavelength bands, the enhancement is by three orders of magnitude. Therefore, relativistic effects can, in principle, provide comprehensive and realistic control over lightsail motion.
\end{abstract}

%\keywords{Suggested keywords}%Use showkeys class option if keyword
                              %display desired

\maketitle

%\tableofcontents

\section{\label{sec:Intro}Introduction}

\begin{figure*}[!htbp]
    \centering
    \includegraphics[width=0.99\linewidth]{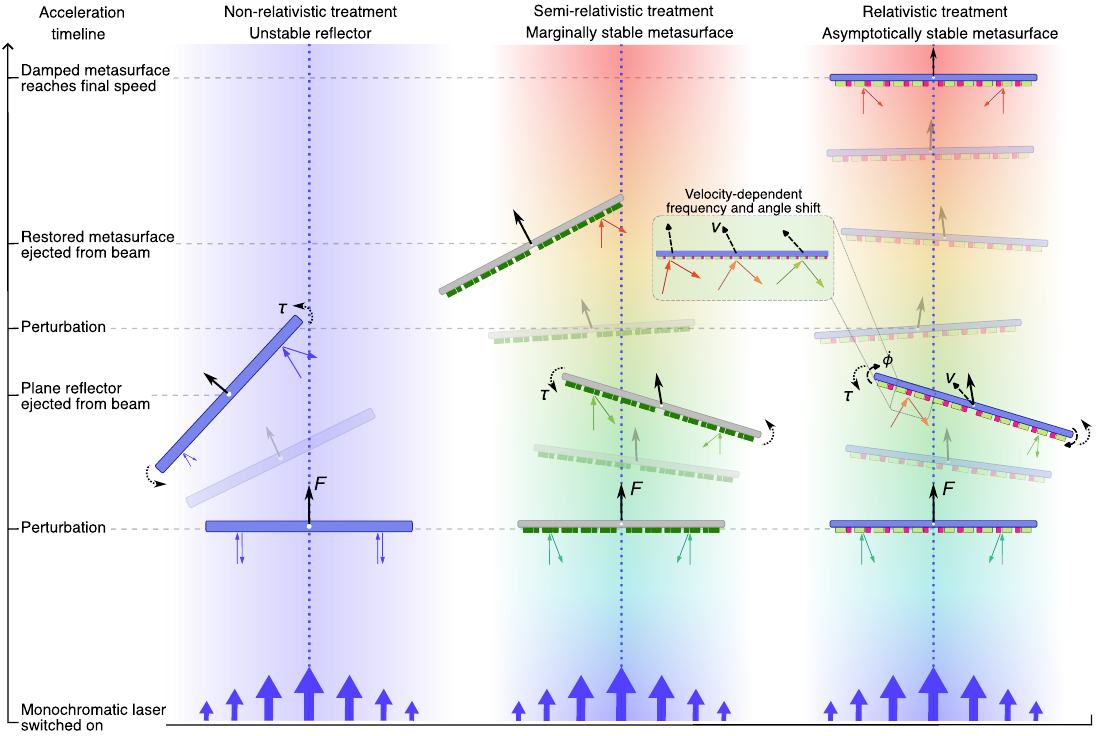}
    \caption{Plane reflectors or designs that are marginally stable (experience restoring mechanisms but not damping) are ejected from the beam due to continual perturbations. Restoring and damping mechanisms arising from a fully relativistic treatment can be harnessed in judiciously patterned membranes, ensuring the lightsail acquires its target velocity.}
    \label{fig:concept}
\end{figure*}

Detailed imaging of neighboring exoplanets such as Proxima B at a distance of 4.2 light years is near-impossible using Earth-based instruments, making long-distance probes an enticing method for determining exoplanet habitability. To approach nearby stars within a human lifetime, probes traveling at near-relativistic speeds (\textit{e.g.} $0.2c$) are necessary. Lightsails are a contender for such missions, with realistic technological progress expected over the next few decades. The current vision considers a payload attached to a \SI{\sim 10}{\square\meter} sail membrane with a small total mass ($<\SI{10}{\gram}$), accelerated by a kilometer-scale, \SI{50}{\giga\watt} laser over several minutes (or over a distance of \SI{0.1}{AU})~\cite{Lubin:2016aa,Lubin:2024aa,Parkin:2024aa}. With a kilometer-scale laser-array aperture, the laser would be capable of focusing on the comparatively narrow meter-scale sail throughout the acceleration phase and thus impart maximum momentum. Critically, the sail must remain within the laser focus. Any perturbations on the sail's motion (\textit{e.g.} due to laser-beam distortion, imperfect tracking or noise) can misalign the sail relative to the beam center, leading to unbalanced forces and torques that generate transverse displacements (relative to the beam propagation direction) and rotations. Due to the large distances involved, feedback to the laser source is too slow, and with no mass budget for active corrective thrusters on the sail, the best option is to use radiation pressure itself to provide stabilization of the sail within the beam. Radiative restoring forces and torques have been proposed, with nanophotonic structures~\cite{Ilic:2019aa,Siegel:2019aa,Srivastava:2019aa,Salary:2020aa,Taghavi:2022aa} emerging as preferred candidates over mirror-based, shaped structures~\cite{Benford:2002aa,Abdallah:2003aa,Manchester:2017aa} in the literature~\cite{Lin:2025aa}. 

However, these restoring mechanisms only provide {\em marginal stability}, meaning restoring forces and torques lead to sail oscillations that do not decay and can in fact grow under stochastic excitation. Marginally stable designs can therefore be perturbed off-course or leave the beam altogether [Fig.~\ref{fig:concept}].   

Thus, in addition to restoring mechanisms, damping mechanisms are necessary. In the right conditions, damping can diminish transverse displacements, rotations and velocities over time, in which case the sail is said to be {\em asymptotically} stable~\cite{Szidarovsky:2017aa} [Fig.~\ref{fig:concept}]. Damping is challenging to implement in the vacuum of space. One proposed damping method is feedback-based parametric stabilization, where the laser is periodically modulated to create time-varying force coefficients~\cite{Chu:2019aa,Chu:2021aa}, reducing the amplitude of the oscillations. Such active-feedback solutions need to be synchronized with the phase of the sail's oscillations, which becomes difficult once the light travel time to the sail greatly exceeds the typical sail oscillation period. 

Slower (adiabatic) time-varying modulation of the restoring forces that do not require synchronization can also reduce oscillation amplitudes~\cite{Salary:2020aa,Salary:2021aa,Taghavi:2022aa}. 
With time-invariance symmetry broken, energy in transverse oscillations is not conserved and can decrease over time~\cite{Landau:1976aa}. This effect was observed in dynamics simulations of dispersive lightsails, with the restoring-force coefficients changing along the sail's trajectory due to the progressive Doppler shift of the laser wavelength. These simulations accounted for the relativistic Doppler effect due to the net center-of-mass motion, but no other relativistic effects, making the analysis ``semi-relativistic'' [Fig.~\ref{fig:concept}].
Moreover, the adiabatic method has yet to show decreasing oscillation amplitudes and velocity amplitudes simultaneously~\cite{Lin:2025aa}. Reducing oscillation velocities is vital for improving the accuracy of the lightsail trajectory once the laser is switched off~\cite{Mackintosh:2024aa}.

A more promising class of passive damping forces and torques is explicitly velocity-dependent mechanisms~\cite{Rafat:2022aa,Mackintosh:2024aa,Lin:2024aa}. Mechanical damping sourced from internal damped modes in sails can strongly diminish oscillations and velocities concurrently~\cite{Rafat:2022aa}, but is challenging to implement in nanometer-thin membranes. Recent work developed a relativistic-wave model to find the optical forces upon a lightsail illuminated by a plane wave~\cite{Lin:2024aa}. The authors found that velocity-dependent relativistic aberration~\cite{Einstein:1905aa} creates damping forces when combined with fine-tuned diffraction gratings. Prior work found that damping torques can be generated by the differential relativistic Doppler shift between different parts of the sail~\cite{Mackintosh:2024aa}, but this effect was only explored for a V-shaped-mirror sail [Fig.~\ref{fig:enhance}(a)]. Such a geometry is unrealistic in practice, but useful to demonstrate the principles of damping and as a damping benchmark. A frequency-dependent optical response was expected to amplify this damping torque, but no detailed investigation was conducted~\cite{Lin:2024aa}. 

In this work, we present the first comprehensive two-dimensional (2D) model for lightsail dynamics in the relativistic regime. The analysis includes the complete set of dynamical effects comprising: the relativistic damping forces and damping torques~\cite{Mackintosh:2024aa,Lin:2024aa}; restoring torques and; restoring forces arising from a finite-width Gaussian laser beam~\cite{Ilic:2019aa,Salary:2020aa}. In this manner, we account for translational and rotational degrees of freedom and their mutual coupling. Moreover, we harness diffraction-grating dispersion to enhance the damping by orders of magnitude over narrow wavelength bands. Thus, we present the first realistic, 2D lightsail analysis with nanostructure-enhanced asymptotic stability embedded passively within the sail design.

We begin in Sec.~\ref{sec:geometry} by defining the laser-lightsail system and highlighting the relativistic damping physics. In Sec.~\ref{sec:Dynamics}, we derive the optical forces and torques for a planar-sail geometry and explain how to integrate the equations of motion accounting for relativity. Then, we discuss linear stability analysis for predicting the stability of sails in Sec.~\ref{sec:Linear Stability Analysis}. Results in Sec.~\ref{sec:results} show sail structures optimized for enhanced asymptotic stability and the resultant damping is showcased in dynamics simulations. Finally, Sec.~\ref{sec:Discussion and Conclusion} discusses the results and their implications.

%%%%%%%%%%%%%%%%%%%%
\section{\label{sec:geometry}Lightsail geometry}

\begin{figure*}[!htbp]
    \centering
    \includegraphics[width=0.99\linewidth]{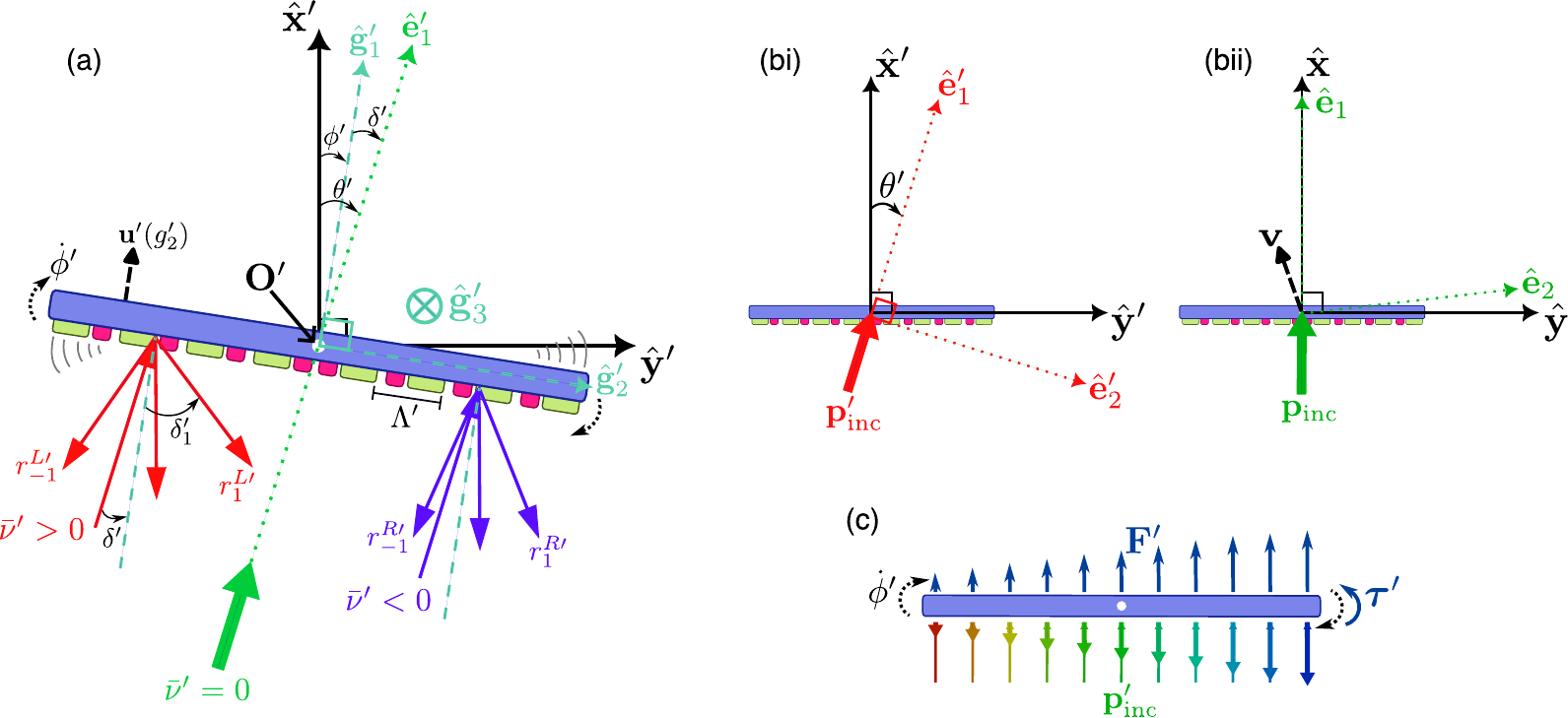}
    \caption{%
    (a) Axes, angles and velocity definitions. The sail depicted here consists of two diffraction gratings connected at the center of mass, $\vb{O}'$. Each grating has a unit cell comprising two rectangular resonators (shown in green and pink). The two gratings are mirror symmetric at $\vb{O}'$ with respect to the normal axis $\hat{\mathbf{g}}_1'$, forming a ``bigrating''.
    (b) Relativistic transformation of light momentum between (i) sail and (ii) laser reference frames. In (i), the incident laser momentum $\mathbf{p}_\text{inc}'$ is relativistically aberrated with non-zero angle $\theta'$ relative to $\hat{\mathbf{x}}'$ and appears Doppler redshifted compared to the laser light in (ii).
    (c) Damping torque on a planar mirror due to position-dependent Doppler shift.
    }
    \label{fig:effects}
\end{figure*}

\subsection{Overview} \label{sec:geometry_overview}
The full description of the geometry and relativistic effects requires multiple reference frames, coordinate systems and a formalism which can rapidly obfuscate the essence of the problem. We thus begin with a simplified overview, entrusting greater detail to subsequent subsections.

We consider a rigid, flat sail in two dimensions, with rest length $L'$ and mass $m$.  Our aim is to calculate the dynamics for a ``bigrating'', where the left and right sections are single diffraction gratings (depicted in Fig.~\ref{fig:effects}(a)), but much of our formalism is more generally applicable to planar sails. The left and right sections must be mirror-symmetric relative to each other at the center-of-mass (CoM) $\mathbf{O}'$, where the sections are attached. Each section reflects light non-specularly, for example in different grating orders, which leads to a net change in momentum components both perpendicular and parallel to the sail. This leads to acceleration (in the direction of the laser) and a transverse force (orthogonal to the laser). 

When the sail is centered on the beam and its plane is aligned orthogonal to the beam axis, the mirror symmetry ensures that the two sections experience identical longitudinal forces and that the transverse forces have opposite directions (and equal magnitude). Thus, there are no net transverse forces and no torques, which defines a dynamic equilibrium position in which the sail is accelerated purely in the direction of the laser, as desired. The two sections are designed so that the net transverse forces and torques are restoring towards equilibrium should the sail be perturbed away from equilibrium (at least for small movements)~\cite{Ilic:2018aa}. 

In the instantaneously comoving (inertial) frame $\frameM$ attached to the CoM, perturbations may rotate the sail. Rotation speeds are non-relativistic so that rotational motion can be calculated using non-relativistic rigid-body mechanics {\em in that frame}. All other kinematics and dynamics need a full relativistic treatment. 

The sail is illuminated by a laser beam modeled as a Gaussian. The laser beam originates from Earth at fixed wavelength $\lzero$, with a wavevector pointing towards the sail's destination. The rapid movement of the sail CoM leads to a longitudinal Doppler shift that can be substantial (up to 22\% for a final velocity of $0.2c$). As the sail accelerates, the laser wavelength measured in frame $\frameM$ is continually redshifting, depicted by the redshift of the green laser between Figs.~\ref{fig:effects}(bi) and~\ref{fig:effects}(bii). Once perturbed, the sail oscillates at relatively low (\textit{i.e.} non-relativistic) velocities in the directions transverse to the beam axis. This slightly modifies the Doppler coefficients compared to a pure longitudinal motion. More significantly, the transverse velocities cause relativistic aberration, a change of the angle of incidence of the light in the sail's reference frame [Fig.~\ref{fig:effects}(b)]. If the sail rotates, each point along the sail has a different velocity relative to the laser. This velocity results in the Doppler shift and relativistic aberration acquiring dependence upon the position along the sail.

\begin{figure*}[!htbp]
    \centering
    \includegraphics[width=0.99\linewidth]{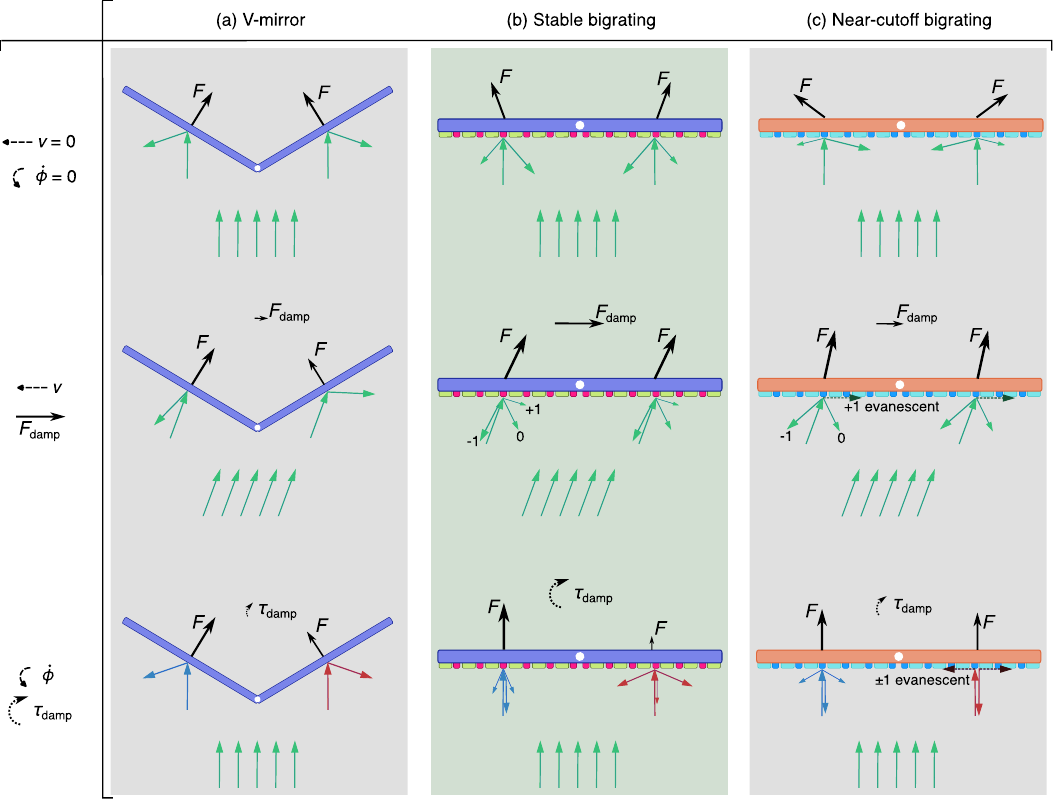}
    \caption{% 
    Relativistic-damping enhancement for sails with translational velocity $v$ or angular velocity $\dot{\phi}$. 
    Column (a): Damping of a V-mirror. 
    Column (b): Dispersive gratings can be damped more strongly than the V-mirror due to having greater control over the scattered light. 
    Column (c): Diffraction gratings operating near cutoff are at risk of losing stability with changes in incident angle or wavelength because non-specular orders responsible for stability may become evanescent.
    }
    \label{fig:enhance}
\end{figure*}

The relativistic aberration and Doppler shift's dependence on position along the sail provide the damping mechanisms~\cite{Mackintosh:2024aa}:
\paragraph{Damping force} Consider a sail with a non-zero velocity transverse to the laser axis. Due to relativistic aberration, the laser light is incident at an angle $\theta'\neq 0 $ that is proportional to the transverse velocity [Fig.~\ref{fig:effects}(b)]. By properly designing the angle dependence of scattered orders, aberration can result in a transverse force proportional to and opposite the transverse velocity~\cite{Lin:2024aa}, \textit{i.e.} a damping force. For plane-mirror sails such as the V-mirror [Fig.~\ref{fig:enhance}(a)], the scattering is highly restricted: small changes in incident angle lead to small changes in forces on the two mirrors. For dispersive sails such as gratings, small changes in angles can lead to a large redistribution of scattered power between the diffraction orders [Fig.~\ref{fig:enhance}(b)]. Through judicious design of the grating geometry, power can be preferentially distributed into the orders responsible for damping forces, thus enhancing the damping compared to plane-mirror designs.
\paragraph{Damping torque} A sail rotating clockwise around its CoM as in Fig.~\ref{fig:effects}(c) has a higher velocity away from the laser on its left tip than on the right tip. Light on the left tip is redshifted relative to the light on the right tip, carrying less momentum. There is thus a force imbalance leading to a torque countering the rotation, \textit{i.e.} a damping torque. The argument is presented for a simple mirror, but as shown in Fig.~\ref{fig:enhance}(b), the damping torque can be enhanced dramatically when reflection coefficients depend on wavelength. With a well-designed grating, small differences in incident wavelength can become large differences in forces between the left and right sides. We will show how to optimize gratings to enhance both damping forces and damping torques in Sec.~\ref{sec:results}.

The scattering properties affect restoring and damping forces and torques, but also lead to cross-coupling, \textit{i.e.} torques due to translation and forces due to rotation. The system is thus rich in dynamics and requires careful optimization to achieve asymptotic stability.

While the relevant effects are perhaps conceptually straightforward, their quantitative analysis requires 
 careful consideration of the dynamics and momentum four-vectors of the light in different reference frames, and appropriate parameterization which we detail below. 

\subsection{\label{subsec:Extension}Frames and coordinates}
We define the bases of the laser-source frame ($\frameL$, assumed to be inertial) and instantaneously comoving (inertial) rest frame ($\frameM$) of the lightsail's CoM as 
\begin{align}
    \vec{f}_0 \xrightarrow[\frameL]{}(1,0,0,0), & & \vec{f}_j\xrightarrow[\frameL]{}(0, \vu{x}_j), \label{eq:basisL} \\
    \vec{f}_{0'} \xrightarrow[\frameM]{}(1,0,0,0), & & \vec{f}_{j'}\xrightarrow[\frameM]{}(0, \vu{x}_{j'}'). \label{eq:basisM}
\end{align}
The four-vector to the left of each arrow has components given by the four-tuple to the right of each arrow, as measured in the frame marked underneath each arrow. For our  2D model, the index $j\in\{1,2\}$ denotes each spatial direction and there is no motion in the third dimension. Primed quantities are measured in $\frameM$, where we define $\vu{x}_{1'}'\equiv\vu{x}'$ and $\vu{x}_{2'}'\equiv\vu{y}'$ for convenience. Similarly, we name the ``longitudinal" direction $\vu{x}_{1}\equiv\vu{x}$ as the sail's intended propulsion direction, and name the ``transverse" (or ``lateral'') direction $\vu{x}_{2}\equiv\vu{y}$. Frame $\frameM$ has velocity $\vb{v} = [v_x,v_y]^T$ with respect to frame $\frameL$, corresponding to the instantaneous velocity of the sail's CoM. The four-basis of $\frameM$ (Eq.~\eqref{eq:basisM}) is obtained from the $\frameL$ basis (Eq.~\eqref{eq:basisL}) via an inverse Lorentz transformation, \textit{i.e.} $\vec{f}_{\mu'}=\Lambda(-\vb{v})^{\nu}_{\mu'} \vec{f}_{\nu}$. Einstein summation is implied over the Greek indices spanning $\{0,3\}$ and the Lorentz transformation matrix $\Lambda(\vb{v})$ is given in Appendix~\ref{app:Lorentz Transformation}. Note $\frameM$ is defined by a Lorentz boost from $\frameL$, which implies both frames' origin coincide at the origin of time. The CoM is thus not generally at the origin of $\frameM$, but at some coordinates $x'(t'),y'(t')$.

We assume a Gaussian laser beam in frame $\frameL$ with intensity $I=I_0 \exp[-2(y/w)^2]$ for $I_0$ the peak intensity and $w$ the beam half-width. For simplicity, we approximate $w$ to be constant along the propagation direction ($\hat{\mathbf{x}}$) in frame $\frameL$. Such a perfectly collimated beam is unrealistic, but in practice could be achieved by dynamically focusing the laser beam to maintain a constant beam width at the sail.
Moreover, since the wavelength of the laser $\lzero$ ($\nu_0 = c/\lzero$ the laser-emission frequency) is much smaller than $w$, we approximate the Gaussian's angular spread to be negligible, characterizing the beam by a single wave four-vector $\vec{k} \xrightarrow[\frameL]{}k_0(1, \vu{x})$, where $k_0=2\pi/\lzero$. 

We define sail-body coordinates $(\gxM, \gyM)$, where $\vu{g}_1'$ is perpendicular to the sail surface and $\vu{g}_2'$ is parallel to the sail surface. The origin of $(\gxM, \gyM)$ coincides with the sail's CoM. For the right (left) sail section, $\gyM \in [0,L'/2]$ ($\gyM \in [-L'/2,0]$), where $L'$ is the length of the full structure. 

To characterize the incident laser momentum, we define light-ray unit vectors $\vu{e}_{1}'$ and $\vu{e}_{2}'$, where $\vu{e}_{1}'$ is parallel to the incident wavevector and $\vu{e}_{2}'$ is the perpendicular unit vector [Fig.~\ref{fig:effects}(bi)]. In general, $\vu{e}_1' \neq \vu{x}'$ due to relativistic aberration as detailed in the next subsection, so we define the angle of incidence $\theta'$ between $\vu{e}_1'$ and $\vu{x}'$.

The sail is considered rigid in $\frameM$, with its orientation characterized by the angle $\phi'$ between the sail normal $\vu{g}_1'$ and $\vu{x}'$. We denote the angle of light incident on $\mathbf{O}'$ with respect to $\vu{g}_1'$ as $\delta'(\vb{v}) = \theta'(\vb{v}) - \phi'$. In $\frameM$, the sail has an angular velocity $\dot{\phi}'$, which must be damped. In our 2D model, the torques are always parallel to the $\vu{g}_3' = \vu{g}_1' \times \vu{g}_2'$ direction.

\subsection{Doppler effect and relativistic aberration}
Since the sail is approaching relativistic speeds, we must take the Lorentz transformation of the light momentum from frame $\frameL$ to frame $\frameM$ in order to accurately describe the relativistic effects in frame $\frameM$. In frame $\frameM$, the incident wavevector has components given by the Lorentz boost $k^{\mu'}=\Lambda(\vb{v})^{\mu'}_{\nu} k^{\nu}$, which introduces the Doppler effect and relativistic aberration~\cite{Einstein:1905aa} responsible for damping. Due to Doppler shift, the wavelength experienced by the sail CoM is $\lambda'(\mathbf{v}) = \lzero/D(\mathbf{v})$, where $D(\vb{v})=\gamma(\vb{v})(1-\beta_x)$ is the relativistic Doppler-factor ($\gamma(\vb{v})=(1-\beta^2)^{-1/2}$ is the Lorentz factor and $\beta=||\vb{v}||/c$ is the normalized sail speed measured in frame $\frameL$). Relativistic aberration appears due to the modest but non-negligible transverse velocities $v_y$, changing the angle of light incident on the sail in frame $\frameM$ as illustrated in Fig.~\ref{fig:effects}(b). Due to relativistic aberration, the light-ray unit vector $\vu{e}_{1}'$ is rotated relative to the $\vu{x}'$ axis, creating the relativistic aberration angle $\theta'$. 

For nonzero angular velocity $\dot{\phi}'$, the magnitudes of the intercepted laser wavelength and relativistic-aberration angle are \textit{not} constant across the surface. This is because the linear velocity of each point on the sail $\vb{u}'(\gyM)$ (normal to the sail surface, see Fig.~\ref{fig:effects}(a)) changes with position $\gyM$ according to $\vb{u}'(\gyM)=-\gyM \dot{\phi}' \vu{g}_1' = -\gyM \dot{\phi}'(\cos\phi' \vu{x}' + \sin\phi' \vu{y}')$. The fact that velocity depends on the surface position means that we must define a new rest frame $\frameM(\gyM)$ at each point across the sail surface that is parametrized by the velocity of the point $\gyM$ in frame $\frameL$. This velocity is given by the relativistic addition $\vb{v}(\gyM) = \vb{v} \oplus \vb{u}'(\gyM)$. In frame $\frameM(\gyM)$, the wavelength of incident light becomes $\lambda'(\gyM)=\lzero/D(\vb{v}(\gyM))$, while the aberration angle $\theta'$ (and by extension $\delta'$) also depends on $\gyM$. The Lorentz transform from $\frameL$ to $\frameM(\gyM)$ is no longer a simple boost, but instead given by a composition of boost and  rotation  $\Lambda(\vb{v} \oplus \vb{u}'(\gyM))R(\varepsilon'(\gyM))$ where  $\varepsilon'(\gyM)$ is the Thomas-Wigner angle~\cite{WignerAngle} (Appendix~\ref{app:Derivation}). 
When expressing angles of aberration using Eq.~\eqref{eq:Aberration} based on a single velocity $\vb{v}(\gyM)$, the Thomas-Wigner rotation of the axes must be taken into account, leading to $\delta'(\gyM) = \delta' - \varepsilon'(\gyM)$. For realistic values of sail rotation velocities, the Wigner correction is 2--3 orders of magnitude smaller than the aberration angle and thus often negligible. However, we keep the Wigner correction in our formalism for completeness.

To quantify the small changes in incident angle and frequency across the surface, we define the angle offset and normalized frequency offset relative to the center of mass as follows:
\begin{align}
\begin{split} \label{eq:offsets}
    \bar{\delta}'(\gyM) &\equiv \delta'(0) - \delta'(\gyM) \,,\\
    \bar{\nu}'(\gyM) &\equiv \frac{\nu'(0) - \nu'(\gyM)}{\nu_0} \,. %\\
    %&\approx -D(\vb{v}) g_2 \cos\phi' \left( \frac{\dot{\phi}'}{c} \right), 
\end{split}
\end{align}

In general, the CoM frame $\frameM$ is not equivalent to $\frameM(\gyM)$ because $\vb{u}'(\gyM)\neq \vb{0}$, and the forces and torques in $\frameM(\gyM)$ must be converted to $\frameM$. The sail rotates slowly such that all parts of the sail have linear velocity $||\vb{u}'(\gyM)|| \ll c$. Hence, the transformation between frames $\frameM(\gyM)$ and $\frameM$ is, to good approximation, Galilean.

The slowly-rotating-sail assumption is valid in the likely regime of the mission according to dynamics simulations that assumed mission parameters consistent with the Starshot project~\cite{Manchester:2017aa,Salary:2020aa,Kumar:2021aa}. In such simulations, typical angular velocities were of order 10--$\SI{100}{\radian\per\second}$, rendering $\dot{\phi}'L$ nonrelativistic. We additionally assume that transverse velocities are nonrelativistic, which is also reasonable given the same simulation results~\cite{Manchester:2017aa,Salary:2020aa,Kumar:2021aa}, where values are $10$--$\SI{100}{\metre\per\second}$ at most. 
Given these nonrelativistic lateral velocities and angular velocities, we derive the equations of motion using Taylor expansions to first order in the small quantities $v_y/c$ and $\dot{\phi}'L/c$.

\section{\label{sec:Dynamics}Dynamics}
\subsection{\label{subsec:Derivation}Equations of motion}
For a non-rotating sail illuminated by a monochromatic plane wave, the force per unit area $s'$ (per unit length in our 2D model) on the sail in $\frameM$ can be expressed as~\cite{Klacka:2008aa}
\begin{align}
    \frac{d\vb{F}'}{ds\,'}&=\frac{D^2 I}{c} \left[ \QprM{1}{\prime} {\vu{e}_{1}}' + \QprM{2}{\prime} {\vu{e}_{2}}' \right]\,, \label{eq:ForcePlane}
\end{align}
where $I$ is the plane-wave intensity.
The efficiency factors $\QprM{j}{\prime}$ quantify the radiation pressure in the direction $\vu{e}_{j}'$ and depend on the scattering properties of the sail, thus depending on wavelength and incident angle of the light. These efficiency factors can be calculated in electromagnetic simulations~\cite{Ilic:2018aa,Lin:2024aa}. If $I$ and $\QprM{j}{\prime}$ depend on position along the sail, as is the case for rotating sails, then provided their spatial variations are on a scale much longer than the wavelength, the net force and torque can be obtained through integration along the sail. 

Projecting the force in Eq.~\eqref{eq:ForcePlane} onto the coordinate axes of frame-$\frameM$  [Fig.~\ref{fig:effects}(a)] and integrating over the surface, we obtain
\begin{widetext}
\begin{subequations}\label{eq:exact_sail_forces}
\begin{align}
    F_{x'}' &= \int_{-L'/2}^{L'/2} d\gyM \, \frac{D^2(\vb{v} \oplus \vb{u}'(\gyM)) I(\gyM)}{c} 
    \left[ 
        \QprM{1}{\prime}(\delta',\lambda') \cos(\theta'+\varepsilon') - \QprM{2}{\prime}(\delta',\lambda') \sin(\theta'+\varepsilon') \right] \,, \\
    F_{y'}' &= \int_{-L'/2}^{L'/2} d\gyM \, \frac{D^2(\vb{v} \oplus \vb{u}'(\gyM)) I(\gyM)}{c} 
    \left[ 
        \QprM{1}{\prime}(\delta',\lambda') \sin(\theta'+\varepsilon') + \QprM{2}{\prime}(\delta',\lambda') \cos(\theta'+\varepsilon') \right] \,, \\
    \bm{\tau}' &= -\vu{g}_3'\int_{-L'/2}^{L'/2} \gyM \, d\gyM \, \frac{D^2(\vb{v} \oplus \vb{u}'(\gyM)) I(\gyM)}{c} 
    \left[ 
        \QprM{1}{\prime}(\delta',\lambda') \cos(\delta'+\varepsilon') - \QprM{2}{\prime}(\delta',\lambda') \sin(\delta'+\varepsilon') \right] \,.
\end{align}
\end{subequations}
\end{widetext}
In these expressions, $\theta'$, $\delta'$, $\varepsilon'$ and $\lambda'$ all depend on $\gyM$.
The factors $\sin(\theta'+\varepsilon')$ and $\cos(\theta'+\varepsilon')$ (and their $\delta'+\varepsilon'$ equivalents) originate from the projection onto the frame $\frameM$ coordinate axes. To lowest order in $\theta'$, these trigonometric factors depend explicitly on sail velocity via relativistic aberration (Appendix~\ref{app:Lorentz Transformation}) and can thus be interpreted as velocity dependence acquired from a changing incident-light angle. The $\QprM{j}{\prime}$ terms, which encode the optical scattering of the sail structure, depend on the incident-light angle and wavelength. Therefore, the relativistic aberration and Doppler effect lead to explicit velocity dependence in the sail's optical response.

A careful analysis of Eq.~\eqref{eq:exact_sail_forces} can reveal the physics anticipated in Section~\ref{sec:geometry_overview}. At normal light incidence on a non-rotating symmetric sail, $\QprM{1}{\prime}$ and $\QprM{2}{\prime}$ are even and odd functions of $\gyM$, respectively. The corrected angle of aberration $\theta'+\varepsilon'$ is zero in the absence of transverse velocity. 
For a sail that is centered on the beam's axis, $I(\gyM)$ is an even function but that becomes untrue once the sail is off-axis. For a sail that is centered on the beam and has zero transverse velocity, the transverse force $F_{y'}'$ thus vanishes (the first term is  $\sin(0)$ and the second term vanishes upon integration). If the sail is off-center ($I(\gyM)$ not an even function), the second term no longer vanishes after integration, providing a restoring force. 
The first term in the integral is proportional to $\sin(\theta'+\varepsilon')$ which comes from the relativistic aberration and is proportional to the transverse velocity $v_y$. Relativistic aberration thus provides a transverse drag force.

For the torque $\bm{\tau}'$, the prefactor in front of the brackets is an odd function of $\gyM$ when the sail is centered and not rotating. Thus, the $\QprM{1}{\prime}$ term integrates to zero and the second term is zero in the absence of transverse velocity, so net torque is zero overall. If the sail is rotating, the Doppler term $D$ is no longer uniform and so the $\QprM{1}{\prime}$ term does not integrate to zero. Therefore, the variable Doppler effect along the sail provides a torque for a rotating sail, which can be used to damp angular oscillations. 
 
For sails with complicated optical responses, integrating the efficiency factors across $\gyM$ is time intensive because the scattering must be sourced numerically from solutions to Maxwell's equations for every value of $\lambda'$ and $\delta'$. Since the rotation of the sail in frame $\frameM$ is nonrelativistic to good approximation, we linearize the sail's optical response to first order in the angular offset and frequency offset around $(\bar{\delta}',\bar{\nu}')=\vb{0}$. For the same reason, we linearize the angles $\theta'(\gyM)$ and $\varepsilon'(\gyM)$ about $\vb{u}'=\vb{0}$ to linear order in $\dot{\phi}' L'/c$. Finally, the intensity distribution $I$ in frame $\frameL$ is evaluated at time $t'$ and position $(x',y') + \gyM(-\sin\phi', \cos\phi')$ in $\frameM$ via the $y$ component of the inverse Lorentz transformation with velocity $\vb{v}$ (since intensity is assumed to only depend on $y$ in $\frameL$). The inclusion of all the effects described in this section, and the resulting equations of motion, are detailed in Appendix~\ref{app:Derivation}. 
 
In Sec.~\ref{sec:Linear Stability Analysis}, further linearization for small movements relative to the dynamic equilibrium provides greater insight into the dominant effects, as well as how to enhance them.

\subsection{Integration}
A significant caveat with the equations of motion (Eqs.~\eqref{eq:exact_sail_forces}) is that they are only valid in frame $\frameM$. Rotations are nonrelativistic in frame $\frameM$, meaning the rotations can be treated in this frame using Newtonian mechanics. 
However, quantities such as the sail's position or velocity must be known and updated in frame $\frameL$. In $\frameL$, the rotating sail cannot be considered rigid because length contractions are different along $\gyM$. That is, no single angle can describe the rotational motion. To resolve this issue, we develop a ``comoving integrator'' for relativistic-lightsail dynamics, which applies rigid-body dynamics in frame $\frameM$ while tracking the sail's position and velocity in frame $\frameL$. 

\begin{figure*}[!htbp]
    \centering
    \includegraphics[width=0.7\linewidth]{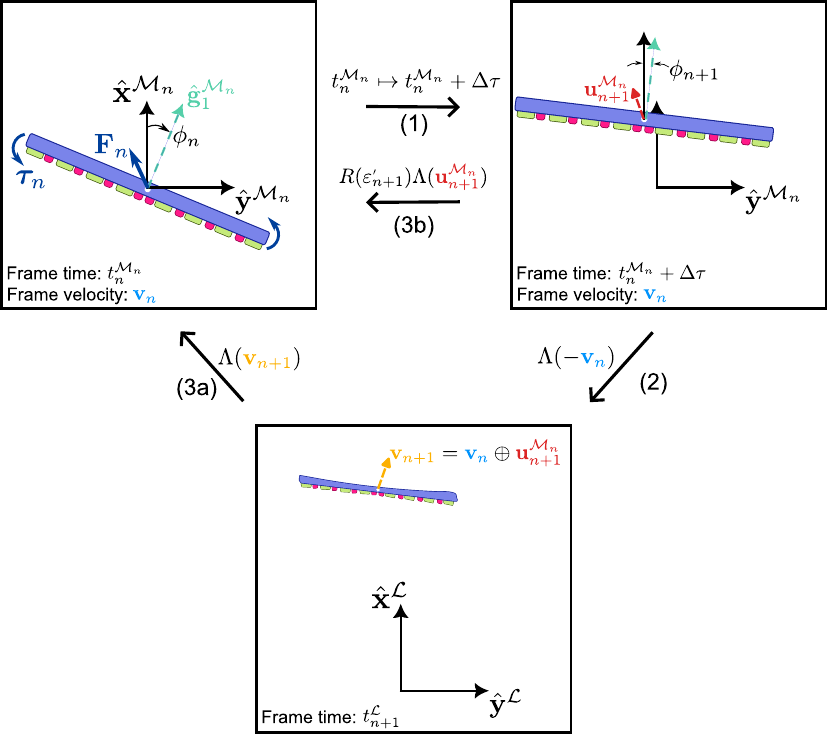}
    \caption{Comoving integration from time step $n$ to time step $n+1$ while recording position and velocity information in frame $\frameL$. In frame $\frameL$, the sail is nonrigid and nonuniformly Lorentz contracted. Note: $\frameM_n$ is defined by a Lorentz boost with no spatial translation, so the spatial origins of frame $\frameM_n$  and frame $\frameL$ coincide at $t^\frameL=t^\frameM=0$. However, for clarity, we have drawn axes where the origin of frame $\frameM_n$ coincides with the sail CoM.}
    \label{fig:comoving}
\end{figure*}

In broad terms, the comoving integrator works as follows. We apply forces and torques in frame $\frameM$ at time $t'$, updating the sail's position, orientation and velocity for time $t'+dt'$. The updated position and velocity information is transferred to frame $\frameL$ via an inverse Lorentz transformation. Then, the sail's new rest frame at $t'+dt'$ is defined by its new velocity in $\frameL$, and the process repeats. Angle and rotation information, which are only meaningful in $\frameM$ due to the sail's non-rigidity in $\frameL$, are not considered in $\frameL$. Therefore, rotational dynamics is only treated in the comoving frame. The method is illustrated in Fig.~\ref{fig:comoving}, which we subsequently describe in detail.

At an arbitrary integration step $n$, the sail is in an instantaneous rest frame $\frameM_n$ at time $t^{\frameM_n}_n$, where the superscript denotes the frame of measurement and the subscript denotes the time step. In $\frameM_n$, the sail's CoM position $(\vb{O}^{\frameM_n}_n$), orientation $(\phi^{\frameM_n}_n)$, angular velocity $(\dot{\phi}^{\frameM_n}_n)$ and linear velocity ($\vb{u}^{\frameM_n}_n = \vb{0}$ by definition of the comoving frame) are known quantities. At step $n$, the sail's CoM moves with velocity $\vb{v}_n$ as measured in frame $\frameL$. After a fixed time step $\Delta \tau$, the new time in $\frameM_n$ is $t^{\frameM_n}_{n+1}=t^{\frameM_n}_{n}+\Delta \tau$, and the lightsail coordinates are evolved using the forces and torques of Eq.~\eqref{eq:exact_sail_forces}. Since the sail's CoM has no velocity in its rest frame, the acceleration $\vb{a}^{\frameM_n}_n$ is \emph{exactly} Newton's second law~\cite{NewtonCorrection}, \textit{i.e.} $\vb{a}^{\frameM_n}_n = \vb{F}^{\frameM_n}_n / m$. For angular coordinates, the rotation-induced linear sail velocity is non-relativistic in $\frameM_n$, justifying a Newtonian approach: $\bm{\alpha}^{\frameM_n}_n = \bm{\tau}^{\frameM_n}_n/J'$, where $J'$ is the moment of inertia. These integrations are indicated by arrow (1) in Fig.~\ref{fig:comoving}, showing the sail's coordinates change to $\vb{O}^{\frameM_n}_{n+1}, \phi^{\frameM_n}_{n+1}, \dot{\phi}^{\frameM_n}_{n+1}$. To first order in the time step, the sail velocity is $\vb{u}^{\frameM_n}_{n+1}\approx \vb{F}^{\frameM_n}_{n} \Delta \tau / m$. We combine the updated time and position of the sail's CoM into a position four-vector:
\begin{align}
    \vec{x}_{n+1} \xrightarrow[\frameM_n]{} 
    \begin{pmatrix}
        t^{\frameM_n}_{n+1}, &\vb{O}^{\frameM_n}_{n+1}
    \end{pmatrix} .
\end{align}
The corresponding time and position of the sail's CoM in frame $\frameL$ comes from the Lorentz boost with velocity $-\vb{v}_n$ of the $\frameM_n$-components of $\vec{x}_{n+1}$. The sail's CoM velocity in $\frameL$ is determined by the relativistic addition ${\vb{v}_{n+1}=\vb{v}_n\oplus\vb{u}^{\frameM_n}_{n+1}}$. The transfer of time, position and velocity information to $\frameL$ is depicted in Fig.~\ref{fig:comoving} by arrow (2). 

Since the sail's CoM has a nonzero velocity in $\frameM_n$ at step $n+1$, frame $\frameM_n$ is no longer a rest frame, so rest-frame forces and torques cannot be applied. Thus, we define the rest frame $\frameM_{n+1}$ as a boost from $\frameL$ with the sail's new velocity $\vb{v}_{n+1}$. The time and position of the sail in $\frameM_{n+1}$ is then given by Lorentz-boosting the frame-$\frameL$ components of $\vec{x}_{n+1}$. This process is depicted by arrow (3a) in Fig.~\ref{fig:comoving} and returns all the necessary information to the sail's updated rest frame except the sail's angular information.

To update angular information between $\frameM_n$ and $\frameM_{n+1}$, we consider the direct transformation between the two inertial frames, depicted by arrow (3b) in Fig.~\ref{fig:comoving}. Frame $\frameM_{n+1}$ is defined by the combination of two Lorentz boosts, which is equivalent to a Wigner rotation and single Lorentz boost $\Lambda(\vb{v}_{n+1})\Lambda(-\vb{v}_{n})=R(\varepsilon_{n+1}')\Lambda(\vb{u}_{n+1})$.
Since $|\vb{u}^{\frameM_n}_{n+1}|\ll c$ we can ignore  deformations in $\frameM_{n+1}$ and treat the sail's orientation in $\frameM_{n+1}$ to simply be rotated by $\varepsilon'_{n+1}$ relative to that in $\frameM_n$, meaning 
\begin{align}
    \phi^{\frameM_{n+1}}_{n+1}&=\phi^{\frameM_n}_{n+1} - \varepsilon'_{n+1}\,.
\end{align}
An identical process is applied to $\dot{\phi}^{\frameM_{n+1}}_{n+1}$, however, it requires evaluating the Wigner-angle derivative $\dot{\varepsilon}_{n+1}'$ numerically using finite differences. The influence of the Thomas-Wigner rotation is largely negligible in any realistic lightsail scenario, with the corrections to angle updates being of order $10^{-4}\%$ at each integration step (see Appendix~\ref{app:Derivation}). Nonetheless, we included the Wigner correction in our code for completeness.

Once angle information has been transferred, the time and state of the lightsail is known in $\frameM_{n+1}$ and the entire process repeats.

%%%%%%%%%%%%%%%%%%%%
\section{\label{sec:Linear Stability Analysis}Linear stability analysis}
The full nonlinear equations of motion derived in Appendix~\ref{app:Derivation} are difficult to interpret. Furthermore, using them to assess sail stability would require a full set of dynamics simulations with different initial conditions.
Instead, we perform a linear stability analysis on the sail in a Gaussian beam, keeping the relativistic corrections. We extract coefficients representing the restoring mechanisms, relativistic damping mechanisms and coupling between rotational and translational degrees of freedom. These enable predictions of asymptotic stability from the eigenvalues of a Jacobian matrix, which we subsequently use to optimize bigrating sails.

Frame $\frameM$ is inappropriate for stability analysis because, by definition, the lightsail's CoM is at rest in this frame. Therefore, we define a new frame $\frameU$ that is centered on the laser-beam axis and has the same longitudinal velocity ($v_x$) as the sail according to $\frameL$. Quantities measured in frame $\frameU$ are marked with two primes. By performing a Lorentz boost with velocity $\vb{v}_x=(v_x,0)$ from $\frameL$ to $\frameU$, the velocity of the sail in $\frameU$ is $(0, \gamma(\vb{v}_x) v_y)$, where $v_y$ is the lateral velocity of the sail in $\frameL$. Hence, frame $\frameU$ and $\frameM$ coincide when $v_y=0$. Since $v_y\ll c$, we approximate that the forces derived in $\frameM$ are valid in $\frameU$, which ceases to be an approximation during linear stability analysis because $v_y = 0$ at dynamic equilibrium. 

We limit the stability analysis to the transverse subspace, \textit{i.e.} without the longitudinal coordinate or velocity.
This is permissible because, to lowest order, near the dynamic equilibrium the longitudinal acceleration decouples from transverse motion and the equations of motion are independent of the $x$ coordinate. Thus, the state vector is $\vb{q}''=[y'', \phi'', v_y'', \dot{\phi}'']$, with desired equilibrium position $\mathbf{q}'' = \mathbf{0}$.
The equation of motion becomes $\dot{\vb{q}}''=\vb{f}''(\vb{q}'')$, where $\vb{f}''(\vb{q}'') = [v_y'',\dot{\phi}'',F_y''/m, \tau'' / (m \Gamma {L''}^2)]$ ($\Gamma=1/12$ in the expression for the moment of inertia of planar sails).  
The forces and torques depend on $v_x/c$ both explicitly, through the relativity factors ($\gamma$ and $D(\mathbf{v})$ in Eq.~\eqref{eq:linear_forces}), and implicitly via the Doppler shift and the sail's dispersive response $\QprM{j}{\prime}(\lambda')$. Therefore, we must treat the longitudinal speed as a system parameter, \textit{i.e.} stability depends on $v_x$.

The left-right symmetry imposed on the sail (and ensuing parity of $\QprM{j}{\prime}$ functions as discussed in Section~\ref{sec:Dynamics}) ensures that forces and torques vanish when $\mathbf{q}'' = \mathbf{0}$. That is, $\mathbf{q}'' = \mathbf{0}$ is an equilibrium point by construction. 
The dynamic stability of the equilibrium is ascertained by linearizing the equations of motion in terms of $\vb{q}''$ about $\vb{q}'' = \vb{0}$, forming a Jacobian matrix 
$[J_{ij}]=[\p f_i'' / \p q_j'']$~\cite{Szidarovsky:2017aa}. 
Denoting the four eigenvalues of $\vb{J}$ as $\xi_i$ ($i=1,2,3,4$), the linearized time evolution is then expressed as a sum of modes with $\exp(\xi_i t'')$ time dependence. The imaginary part of $\xi_i$ represents oscillatory motion (due to restoring forces/torques) and the real part describes exponential growth or decay. 
Asymptotic stability occurs when all four eigenvalues of $\vb{J}$ have negative real parts, that is,  $\Re(\xi_i)<0$. 
Since the Jacobian coefficients are real, the Jacobian eigenvalues are either real or complex-conjugate pairs. Purely real eigenvalues imply no restoring behavior, which is undesirable, so we seek sails with two complex-conjugate eigenvalue pairs.

We express $\vb{J}$ as 
\begin{equation} \label{eq:JacobianMatrix}
    \vb{J}
    =
    \begin{pmatrix}
        0 & 0 & 1 & 0 \\
        0 & 0 & 0 & 1 \\
        \kyy & \kyp & \myy & \myp \\
        \kpy & \kpp & \mpy & \mpp
    \end{pmatrix}\,. 
\end{equation}

The ``restoring'' terms are force derivatives with respect to $y''$ or $\phi''$ (labeled by $k$), whereas the ``damping'' terms are force derivatives with respect to $v_y''$ or $\dot{\phi}''$ (labeled by $\mu$). The terms $\kyy$ and $\kpp$ are the restoring force and torque coefficients, respectively, while $\myy$ and $\mpp$ are the damping force and torque coefficients, respectively. Intuitively, we expect that asymptotic stability is obtained when the restoring force and torque coefficients have negative sign, which should also be true of the damping coefficients. However, more generally, asymptotic stability depends on the nontrivial combination of the Jacobian coefficients into the eigenvalues $\xi_i$, which includes the terms that encode coupling between rotational and translational degrees of freedom ($\kyp$, $\kpy$, $\myp$ and $\mpy$). Observe that there are no {\em a priori} symmetries in the Jacobian: there is generally no relation between $\kpy$ and $\kyp$, or $\mpy$ and $\myp$.
With the exception of Ref.~\cite{Rafat:2022aa}, past linear stability analyses of lightsails have only included {\em ad hoc} damping terms without justifying their physical origin~\cite{Ilic:2019aa,Srivastava:2019aa}. Moreover, the coupled-damping terms $\myp$ and $\mpy$ have consistently been set to zero when acknowledged at all~\cite{Ilic:2019aa, Shirin:2021aa}. 

The eight nontrivial Jacobian terms are found through linearization of Eq.~\eqref{eq:linear_forces} about $\vb{q}''=\vb{0}$ and are listed in Appendix~\ref{app:Linear Stability}. Among them, we highlight the damping force and damping torque as follows:
\begin{align}
    \myy(\lambda')
    &=
    - D^2 \frac{2P_0}{mc} \frac{1}{c} \frac{D+1}{D(\gamma+1)} \ws{1/2} 
    \left[ 
        \QprM{1}{R\prime} + \frac{\p \QprM{2}{R\prime}}{\p \delta'} 
    \right]\! (\lambda') 
    \,, \label{eq:mu_yy}
    \\
    \mpp(\lambda')
    &=
    - D^2 \frac{2P_0}{mc} \frac{1}{c} \frac{1}{\Gamma} \wf{0} 
    \left[
        2 \QprM{1}{R\prime} - D \frac{\p  \QprM{1}{R\prime}}{\p {\bar{\nu}'}} 
    \right]\! (\lambda')
    \,. \label{eq:mu_phiphi}
\end{align}
The factor 2 multiplying the laser power $P_0$ comes from the constructive sum of efficiency factors on the left and right sail halves such that only the right-half efficiency factors $\QprM{1}{R\prime}$ and $\QprM{2}{R\prime}$ appear. The factor $1/c$ is only present for the damping terms of the Jacobian, making them of order $10^{9}$ times smaller than the restoring terms. However, the damping terms can be preferentially enhanced by the optical response of the sail, which is encoded in their dependence on the efficiency factor derivatives. The effect of the Gaussian intensity profile is manifest in the $\ws{1/2}$ and $\wf{0}$ terms, which have a nontrivial dependence on the Gaussian-beam width (see Appendix~\ref{app:Linear Stability}).

The damping force and torque coefficients in Eqs.~\eqref{eq:mu_yy} and~\eqref{eq:mu_phiphi} have two contributions. Common to both coefficients is the $\QprM{1}{\prime}$ term, which is the only surviving contribution if the sail has no frequency or angle dispersion. This contribution is weak because it is bounded by $\QprM{1}{\prime}(\delta'=0) = 2$, which is the well-known factor of 2 that appears in the radiation pressure on a mirror reflecting 100\% of incident power normal to its surface. In contrast, the angular dispersion ($\p\QprM{2}{\prime}/\p\delta'$)~\cite{Lin:2024aa} and frequency dispersion ($\p\QprM{1}{\prime}/\p\bar{\nu}'$) terms can be enhanced for substantially stronger damping. 
These dispersive terms depend on the wavelength of incident light, which in turn varies with the velocity $v_x$ of the sail due to the relativistic Doppler effect. 

For a laser with fixed Gaussian width and a sail with known longitudinal velocity, the presence (and degree) of asymptotic stability is generally deduced by numerically calculating the Jacobian eigenvalues. Since the $\mu$ are small, we can acquire analytic insight using first-order perturbation theory (derived in Appendix~\ref{app:perturbation}), yielding
\begin{widetext}
\begin{align} \label{eq:pert_eigval_jac}
    \xi_{1,2}
    &=
    \xi_{\text{unp},1,2} +
    \frac{1}{4} 
    \bigg[ 
        \myy + \mpp - \frac{2(\kyp\mpy + \kpy\myp) + (\kyy - \kpp)(\myy - \mpp)}{\sqrt{(\kyy - \kpp)^2 + 4\kpy\kyp}}
    \bigg] 
    \,, \\
    \xi_{3,4}
    &=
    \xi_{\text{unp},3,4} +
    \frac{1}{4} 
    \bigg[ 
        \myy + \mpp + \frac{2(\kyp\mpy + \kpy\myp) + (\kyy - \kpp)(\myy - \mpp)}{\sqrt{(\kyy - \kpp)^2 + 4\kpy\kyp}}
    \bigg] 
    \,,
\end{align}
\end{widetext}
which is in good agreement with numerically calculated eigenvalues. In these equations, the unperturbed terms $\xi_{\text{unp}}$ can often be interpreted as the imaginary parts of the eigenvalues, while the perturbation contributions are the real parts (Appendix~\ref{app:perturbation}). Therefore, asymptotic stability appears to benefit from maximizing the damping force and torque coefficients and minimizing the cross-coupling terms. However, we cannot make the rotation-translation coupling terms disappear, because they depend on the same optical coefficients as the pure restoring or damping terms (Eqs.~\eqref{eq:Jacobian Terms}). For instance, we calculate
\begin{equation} \label{eq:mpy}
    \mpy(\lambda') 
    = 
    \frac{1}{c} \frac{D+1}{D(\gamma+1)} \kpp(\lambda') 
    \,,
\end{equation}
which cannot be zero because stability requires a non-zero restoring torque ($\kpp \ne 0$). There are similar relations between the other cross-coupling coefficients. Hence, the cross-coupling, damping and restoring mechanisms are inseparable in the stability of lightsails via passive optical scattering.

\section{\label{sec:results}Results}
In this section, we apply the linear stability analysis to optimize for asymptotic stability in a particular sail structure and showcase the resultant bounded and decaying dynamics. Following previous analyses of marginally stable designs~\cite{Ilic:2019aa} and designs with purely translational damping forces~\cite{Lin:2024aa}, we choose a ``bigrating'' sail, comprised of a diffraction grating connected on one end to its mirror-symmetric counterpart [Fig.~\ref{fig:effects}(a)]. As in Ref.~\cite{Lin:2024aa}, we assume a perfectly rigid sail that scatters light into the specular and first diffraction orders only, which is easily generalizable to higher orders. We will treat cases with transmission and without transmission separately. We simulate gratings using rigorous coupled-wave analysis assuming TE (out-of-plane) polarization, but the results can be extended to other polarizations. The reflected/transmitted orders are well approximated as rays with (normalized) power efficiencies $[r,t]_m'(\delta',\lambda')$ (for $m=0,\pm1$) and angles $\delta_m'$ that obey the grating equation: $\sin\delta_m' = \sin\delta' + m\lambda'/\Lambda'$ [Fig.~\ref{fig:effects}(a)]. The efficiency factors $\QprM{j}{\prime}$ can be expressed in closed form by applying momentum conservation on these scattering efficiencies~\cite{Lin:2024aa} (Appendix~\ref{app:grating}), with the resulting forces obtained by the method described in Sec.~\ref{sec:Dynamics}.

To design an asymptotically stable bigrating, we employ inverse design based on previous optimization strategies~\cite{Jin:2020aa,Lin:2024aa}. The objective function is
\begin{equation} \label{eq:FOM}
    F_\text{asymp} 
    = 
    % \braket{\xi_\text{max}(\lambda')}_{\lambda'} 
    % = 
    \Bigg\langle
        \text{maximum}[\Re(\bm{\xi})]
    \Bigg\rangle_{\lambda'} \,,
\end{equation}
which is an average (denoted by angle brackets) over the Doppler-shifted wavelength range $[\lzero,\lmax]$, corresponding to the sail's acceleration from $v=0$ to $v=v_f$ ($\lmax\equiv\lzero/D(\vb{v}_f)$). The vector $\bm{\xi}$ contains all eigenvalues of the Jacobian. Minimizing the objective function corresponds to minimizing the real part of the eigenvalue with the largest real part, which drives all eigenvalues to having negative real parts with large magnitude. We refer to the two eigenvalues containing the degenerate larger-magnitude negative real-part as the ``weak'' eigenvalues and the two eigenvalues with the degenerate smaller-magnitude negative real part as the ``dominant'' eigenvalues. The naming corresponds to the modes $e^{\xi_j t''}$ that are weak/dominant in magnitude after long times. 

\begin{figure*}[!htbp]
    \centering
    \includegraphics[width=0.85\linewidth]{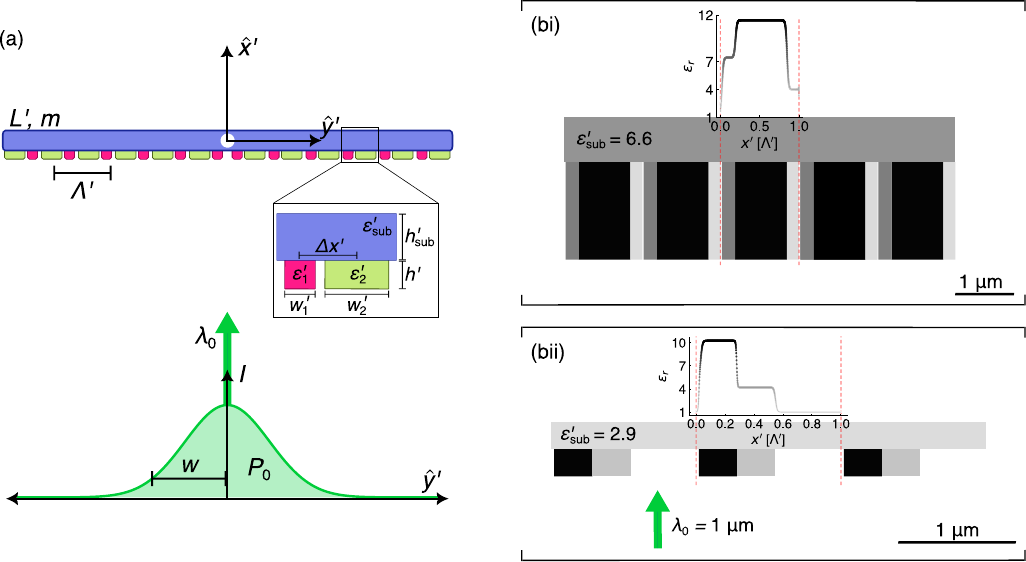}
    \caption{%
    (a) Bigrating sail, unit cell and optimization parameters.
    (b) To-scale schematics of the (i) narrow-band [Sec.~\ref{sec:narrow_band}] and (ii) broadband optimized gratings [Sec.~\ref{sec:broadband_grating}] (exact parameters in Appendix~\ref{app:Numerical optimisation}). The unit cells are shown between the dashed red lines. The gray level of the substrate and resonators indicates their permittivity in accordance with the plotted permittivity profiles. Note that different scales are used in (bi) and (bii).}
    \label{fig:structure}
\end{figure*}

We set nominal total sail mass $m$, laser output power $P_0$ and surface area $L'$ according to recent estimates~\cite{Lubin:2016aa} as $\SI{1}{\gram}$, $\SI{50}{\giga\watt}$ and $\SI{10}{\meter}$, respectively. 
We assume a Gaussian-beam width of $w = 2L'=\SI{20}{\metre}$, which is discussed in Appendix~\ref{app:Numerical optimisation}. 
We consider a diffraction grating with a unit cell comprised of two different dielectric resonators [Fig.~\ref{fig:structure}], inspired by previous designs~\cite{Ilic:2019aa}. The optimization procedure searches a parameter space consisting of, at most: the grating period ($\Lambda'$) and grating-layer thickness ($h'$); the resonator widths ($w_1'$, $w_2'$), permittivities ($\epsilon_{1}'$, $\epsilon_{2}'$) and their central separation ($\Delta x'$); the substrate thickness ($h_\text{sub}'$) and permittivity ($\epsilon_\text{sub}'$). 
For further details on our optimization procedure, see Appendix~\ref{app:Numerical optimisation}.

\subsection{Reflection-only gratings}
We begin by studying the simplest case of purely reflecting gratings, which we simulate by taking the substrate as a lossless, reflecting material (negative, large-magnitude permittivity). In this case, applying symmetry, energy conservation and reciprocity~\cite{Petit:1980aa,Loewen:2017aa} simplifies the Jacobian coefficients into containing just two independent reflection efficiencies and their wavelength and angle derivatives (see Appendix~\ref{app:grating}). Of these Jacobian coefficients, we highlight the restoring force and restoring torque terms:
\begin{align}
    \kyy(\lambda')
    &=
    -D^2 \frac{2P_0}{mc} \frac{1}{L'} \winf [\refl{R}{-1} - \refl{R}{1}] \sin\delta_1' \,,
    \\
    \kpp(\lambda')
    &=
    - \frac{1}{\Gamma} \frac{1}{\cos\delta_1'} \kyy(\lambda') \frac{\ws{0}}{\winf} 
    \propto -\kyy(\lambda')
    \,.
\end{align}
The terms $\ws{0}$ and $\winf$ come from the Gaussian-beam width dependence and are strictly positive. We observe that $\kpp$ necessarily has the opposite sign to $\kyy$, meaning that purely reflecting gratings cannot have both restoring forces \emph{and} restoring torques simultaneously. Therefore, achieving asymptotic stability for such gratings relies on engineering the rotation-translation coupling terms (see Eqs.~\eqref{eq:Jacobian Terms}) rather than the restoring force and torque. 
Since perfectly-reflecting gratings are advantageous for lightsail propulsion, the anti-restoring behavior demonstrated here complicates future optimizations that seek dynamical stability and maximal propulsion concurrently. 
However, the anti-restoring response does not apply to purely-reflecting gratings that diffract into higher order modes ($|m|>1$), nor other structures such as aperiodic reflective metasurfaces.

\subsection{Narrow-band reflection-transmission gratings\label{sec:narrow_band}}
\begin{figure*}[t!]
    \centering
    \includegraphics[width=0.95\textwidth]{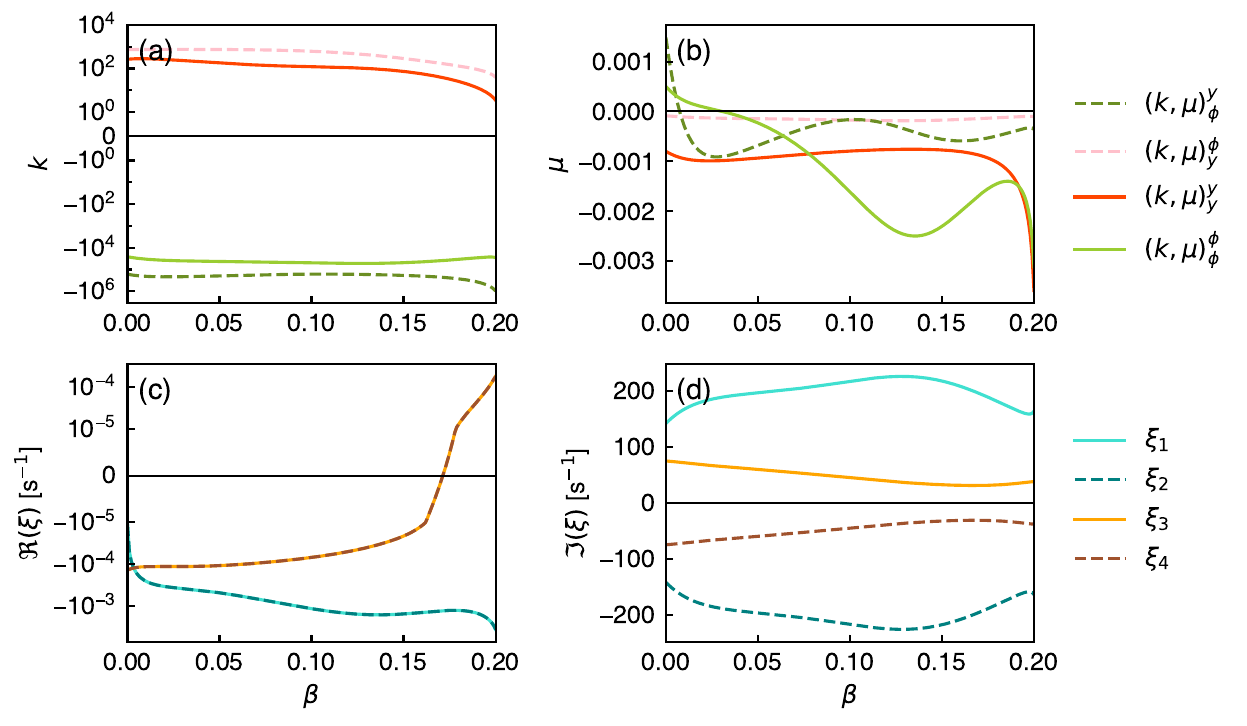}
    \caption{(a) Restoring coefficients, (b) damping coefficients, (c) real-part of eigenvalues and (d) imaginary-part of eigenvalues over velocity/wavelength for the broadband-optimized bigrating shown in Fig.~\ref{fig:structure}(bii). The units of the restoring terms are: $[\kyp] = \SI{}{\meter\per\square\second\per\radian}$, $[\kpy] = \SI{}{\radian\per\meter\per\square\second}$, $[\kyy] = [\kpp] = \SI{}{\per\square\second}$, while the corresponding damping terms have the same units multiplied by \SI{1}{\second}. 
    }
    \label{fig:eigenvalues}
\end{figure*}

For the remainder of this section, we turn to gratings with transmissive dielectric substrates whose thickness and permittivity are variables in the optimization. To gauge the possible damping we can achieve over broad wavelength ranges, we start by minimizing Eq.~\eqref{eq:FOM} at a single wavelength ($\lzero = \SI{1}{\micro\meter}$), searching for narrow-band resonant enhancement. 
Narrow-band effects can be harnessed in nonrelativistic lightsails that can explore the solar system within days~\cite{Santi:2023aa,Parkin:2024aa}, since smaller Doppler shifts correspond to smaller bandwidths. For this single-wavelength optimization, we allowed the grating pitch to vary between that associated with the first- and second-order diffraction cutoffs. 

After searching the parameter space, the optimizer discovered the grating shown in Fig.~\ref{fig:structure}(bi), whose eigenvalues have real parts around $\Re(\xi_\text{dominant}) \sim \SI{-6e-3}{\per\second}$ and $\Re(\xi_\text{weak}) \sim \SI{-0.4}{\per\second}$ over a bandwidth of just \SI{40}{\pico\meter}. Thus, this grating relies on a sharp resonance to enhance the damping (sensitivity to parameter variations is discussed in Appendix~\ref{app:Numerical optimisation}).
The optimized grating pitch $\Lambda' = 1.328\lzero$ is well above the first-order cutoff, indicating that the resonance comes from dispersion enhancement that does not rely on grating-cutoff effects. The optimized real parts correspond to characteristic damping timescales of around \SI{170}{\second} and \SI{3}{\second}, respectively, in comparison to the Starshot acceleration timescale of around \SI{1000}{\second} for a $0.2c$ target. Therefore, if eigenvalues with these magnitudes could be sustained over a broad wavelength band, then relativistic optical damping would be a highly potent stabilization mechanism. 

We cannot directly compare the dynamics performance of our optimized-grating design with that of mirror-based sails like the V-mirror~\cite{Mackintosh:2024aa} because mirror-based sails typically require hollow (not Gaussian) laser-beam intensity profiles for stable beam riding. Therefore, we compare the linear damping coefficients derived here for a dispersion-enhanced grating (Eqs.~\eqref{eq:Jacobian Terms}) with those of the V-mirror derived from the equations of motion of previous work~\cite{Mackintosh:2024aa,Lin:2024aa} (which were derived assuming a plane-wave laser). For fair comparison, both sails intercept the same laser power ($P_\text{intercept}=\SI{19}{\giga\watt}$) and have the same mass ($m=\SI{1}{\gram}$). Additionally, we select the best V-mirror for the damping comparison, which maximizes the product $\myy \mpp$ at a mirror angle $\alpha_0' = \SI{45}{\degree}$~\cite{Mackintosh:2024aa}. The V-mirror values are $\braket{\myy}_{\lambda'} = \braket{\mpp}_{\lambda'} \approx -\SI{4.3e-4}{\per\second}$ (the average is taken over a \SI{40}{\pico\meter} bandwidth, where the damping coefficients are effectively constant). For our narrow-band optimized grating, the translational and rotational damping coefficients averaged over the \SI{40}{\pico\meter} bandwidth are $\braket{\myy}_{\lambda'} = \SI{-3.5e-4}{\per\second}$ and $\braket{\mpp}_{\lambda'} = \SI{-0.93}{\per\second}$, respectively. The translational damping is slightly worse than the optimal V-mirror, but the rotational damping coefficient is enhanced by a factor 2000, showcasing the potency of dispersion enhancement. To see if such enhancements are possible over broader bands ($>\SI{200}{\nano\meter}$) in $v_f=0.2c$ missions, we conduct a broadband optimization in Sec.~\ref{sec:broadband_grating}.

\subsection{Broadband reflection-transmission gratings\label{sec:broadband_grating}}
For broadband gratings, the optimization is conducted using Eq.~\eqref{eq:FOM}, including the average over the complete wavelength range associated with acceleration to $0.2c$. In contrast to the single-wavelength optimization, we fixed the grating pitch to a value slightly larger than $\lmax$ (see Appendix~\ref{app:Numerical optimisation}). The danger of setting grating pitch too close to cutoff is that it reduces the range of incident angles for which the $m=\pm1$ orders propagate. These non-specular orders are necessary for restoring and damping forces to appear. Therefore, during optimization, we fixed the grating pitch such that the $m=\pm1$ orders are propagating in the range of incident angles $\delta'\in[-\SI{0.1}{\degree},\SI{0.1}{\degree}]$, even up to the maximum wavelength $\lambda' = \lmax$. 

The best design found across several optimizations is shown schematically in Fig.~\ref{fig:structure}(bii), with the associated grating parameters tabulated in Appendix~\ref{app:Numerical optimisation}.
Figure~\ref{fig:eigenvalues} shows the Jacobian coefficients and eigenvalues for this grating. 
The damping coefficients $\myy$ and $\mpp$ in Fig.~\ref{fig:eigenvalues}(b) tend to larger-magnitude negative values as $\beta\rightarrow0.2$, showing that the grating utilizes dispersion enhancement near the first-order cutoff. Interestingly, the grating has an anti-restoring force (solid orange curve $\kyy$ in Fig.~\ref{fig:eigenvalues}(a) is above zero). Asymptotic stability in this grating is therefore achieved by balancing the sign and magnitude of the cross-coupling terms. Unlike purely-reflecting gratings where the anti-restoring response is unavoidable, transmissive gratings can provide restoring forces and torques simultaneously. The optimizer discovered such gratings, but their real-part eigenvalues averaged over wavelength are smaller than those in Fig.~\ref{fig:eigenvalues}.

The average damping coefficients of the grating in Fig.~\ref{fig:eigenvalues} are $\braket{\mu_y^y}_{\lambda'} \approx -\SI{9.4e-4}{\per\second}$ and $\braket{\mu_\phi^\phi}_{\lambda'} \approx -\SI{1.2e-3}{\per\second}$. For the V-mirror, the damping coefficients averaged over the $0.2c$-Doppler band are $\braket{\myy}_{\lambda'} = \braket{\mpp}_{\lambda'} \approx -\SI{3.4e-4}{\per\second}$. Thus, over a wavelength range corresponding to $v_f = 0.2c$, we see almost 3 times improvement in the translational damping coefficient and almost 4 times improvement in the rotational damping coefficient compared to the best V-mirror sail. Therefore, even over a broad wavelength range, damping enhancement is possible through nanostructured-dispersion design. The average real-part of the eigenvalues over wavelength are of order $\braket{\Re(\xi_\text{dominant})}_{\lambda'} \sim \SI{-e-5}{\per\second}$ and $\braket{\Re(\xi_\text{weak})}_{\lambda'} \sim \SI{-e-2}{\per\second}$. 
The dominant mode has a long characteristic decay timescale of around 30~hours, whereas the weak mode decays over a timescale of \SI{100}{\second}.
Relative to the acceleration period of around 10~minutes, the weak mode is strongly attenuated. The dominant mode is only mildly damped, but does not grow on average over the acceleration phase, which may not be the case for sails optimized purely for marginal stability~\cite{Ilic:2019aa}. Although the average of the real-part eigenvalues over wavelength is negative as prescribed by Eq.~\eqref{eq:FOM}, there may be wavelength ranges where the real part becomes positive. For instance, $\Re(\xi_\text{dominant})$ becomes positive beyond $\beta = 0.17$ in Fig.~\ref{fig:eigenvalues}(c). In general, these growth regions endanger beam-riding stability, especially over long times or when the sail encounters sustained perturbations. However, in this case, the dominant-eigenvalue real part is 10 times smaller than the weak-eigenvalue real part (note the log scale) and corresponds to a growth timescale of at least 3~hours. Given that the growth for this optimized grating only occurs in the final 15\% of the acceleration phase (minutes long at most), the effect on stability is minimal. We have found gratings with negative real-part eigenvalues over the entire velocity range, but their average real-part eigenvalue magnitudes were smaller than those in Fig.~\ref{fig:eigenvalues}(c).

\begin{figure*}[!htb]
    \centering
    \includegraphics[width=0.95\textwidth]{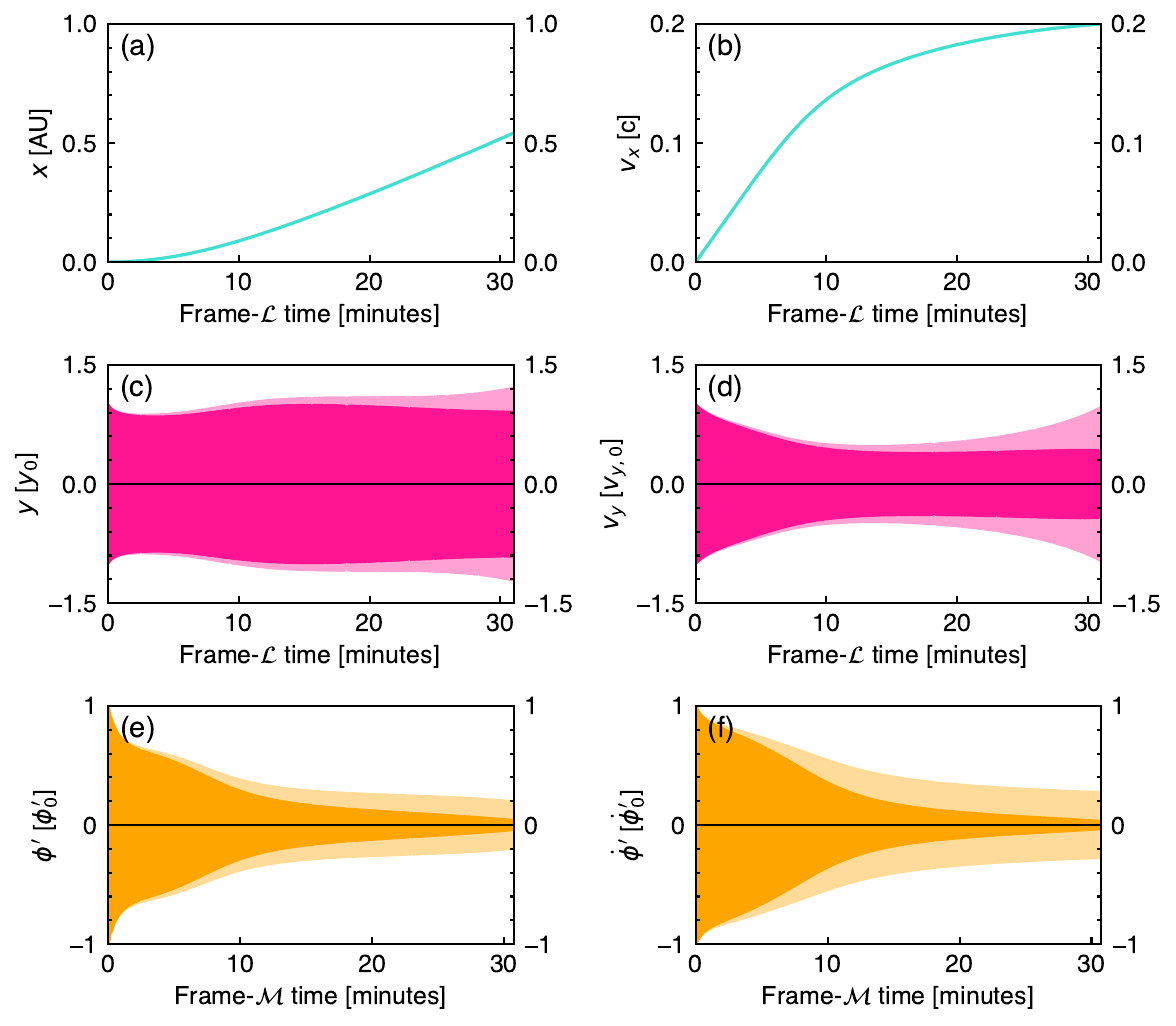}
    \caption{Optimized-bigrating [Fig.~\ref{fig:eigenvalues}] dynamics. Panels~(a) and~(b) show the longitudinal propulsion until the sail reaches $v_f=0.2c$, and panels~(c)--(f) show the transverse coordinates. In the latter panels, the nonlinear dynamics with the damping terms enabled are colored deeply, whereas the linear dynamics with the damping terms disabled are colored lightly. Each transverse coordinate is displayed in units of its initial envelope value.}
    \label{fig:dynamics}
\end{figure*}

The eigenvalues and mode-decay timescales just discussed only estimate the damping performance. A more complete assessment involves a dynamics simulation, which accounts for the exact acceleration time during which the damping occurs. For the grating in Fig.~\ref{fig:structure}(bii), we integrated the equations of motion using the solver developed in Sec.~\ref{sec:Dynamics}. With the initial condition $[x_0,y_0,\phi'_0,v_{x,0},v_{y,0},\dot{\phi}'_0] = [0, 0.001w,\SI{0.01}{\degree},0, -\SI{0.5}{\meter\per\second}, -\SI{0.005}{\rps}]$, the resultant dynamics is displayed in Fig.~\ref{fig:dynamics}. In these plots, the oscillations around equilibrium are not visible because of the extremely short period (of order \SI{0.1}{\second}) relative to the total flight time. Due to the Doppler shift, the wavelength and hence $\QprM{j}{\prime}$ and the force coefficients vary over time, leading to the time-varying envelopes in the dynamics. Even in the absence of damping, the restoring ($k$) terms retain this implicit time dependence, which contributes heavily to the envelope shape. This effect is pronounced in dispersive sails such as gratings. To quantify the damping effectiveness, we display the coordinates in units of their initial envelope value, rather than the exact initial condition. Figures~\ref{fig:dynamics}(c--f) show the integrated nonlinear dynamics (Eq.~\eqref{eq:linear_forces}, dark color) and the integrated linear dynamics (Eq.~\eqref{eq:linear_eom}, light color), with the latter having $\mu^i_j$ ($i,j\in\{y,\phi\}$) artificially set to zero. 
When the damping is disabled, the adiabatic changes in restoring coefficients lead to final values of $[y,v_y,\phi',\dot{\phi}'] = [120\%,19\%,96\%, 27\%]$ relative to the initial envelope value. With damping included in the simulations, these are substantially reduced to $[y,v_y,\phi',\dot{\phi}'] = [91\%,4\%,42\%, 3\%]$. Therefore, Fig.~\ref{fig:dynamics} shows asymptotic stability of a lightsail achieved via relativistic damping, with significant reductions in both rotational and translational coordinates and velocities. 
We highlight that, even in the absence of damping, the dispersive torques are responsible for substantially reducing $\phi'$ and $\dot{\phi}'$ simultaneously in Figs.~\ref{fig:dynamics}(e) and~(f). However, such adiabatic effects have not yet proven capable of diminishing oscillation amplitudes and velocities simultaneously across both rotation and translation degrees of freedom (see Sec.~7 of Ref.~\cite{Lin:2025aa}), evidenced by the noticeable increases in $y$ and $v_y$ when damping is disengaged. In Fig.~\ref{fig:dynamics}, explicit velocity-dependent damping is shown to diminish all transverse degrees of freedom simultaneously. Regardless, such strong adiabatic responses warrant further study in tandem with the damping effects showcased here. 

Aside from the result in Fig.~\ref{fig:dynamics}, there are other initial conditions for which the broadband grating is dynamically stable. In general, it is difficult to determine the stable region of phase space. As a rough estimate of this stable region, we repeated the dynamics simulation for a variety of initial values (see Appendix~\ref{app:Dynamics} for details). Taking only a single coordinate per perturbation to be initially perturbed, we found that the sail reaches $v_f=0.2c$ for magnitudes up to $y_0 = 0.06w$, $\phi_0' = \ang{8}$, $v_{y,0} = \SI{3}{\meter\per\second}$ and $\dot{\phi}'_0 = \SI{1}{\rps}$, which are fairly large initial offsets.
In some of these simulations, the $m=\pm1$ orders become evanescent, especially at higher velocities where the Doppler shift is larger. This is because the range of rotation angles $\phi'$ for which the $m=\pm1$ diffracted orders are propagating progressively decreases as the sail approaches $v_f=0.2c$ (according to the grating equation). Provided that $\phi'$ only slightly exceeds the cutoff angle, only one of $m=1$ or $m=-1$ are cutoff and the sail can still reach its target velocity [Fig.~\ref{fig:enhance}(c)]. However, allowing orders to become evanescent is dangerous if the sail experiences large and/or continual perturbations, which is likely the case in the Starshot mission.

%%%%%%%%%%%%%%%%%%%%
\section{\label{sec:Discussion and Conclusion}Discussion and conclusions}
In this paper, relativistic damping was shown to be effective at providing asymptotic stability for laser-driven lightsails when harnessing grating resonances. 
However, the degree of damping enhancement with nanophotonic structures can, in principle, be increased further. Improvements may come from alternative optimizations (\textit{e.g.} over a larger parameter space with a neural-network-enhanced search~\cite{Norder:2025aa}) or a multiple-grating design~\cite{Ilic:2019aa,Taghavi:2022aa}. Multiple gratings can divide the wide Doppler band by having each grating operate at a preset portion of the full wavelength range. With narrow-band resonances, the dispersion enhancement within the operating band of the individual gratings can be extreme; comparing the optimal eigenvalues of Sec.~\ref{sec:narrow_band} and Sec.~\ref{sec:broadband_grating}, the best eigenvalues over narrow bands were 10--100 times larger than the best eigenvalues over the full Doppler spectrum. By combining multiple gratings that contain such enhancement, we expect broadband damping that is orders of magnitude larger than that obtained with the single grating. The main issue with the multiple-grating approach is the possibility of the out-of-band, ``inactive" gratings generating instability.

The degree of damping enhancement provided by dispersive membranes clearly depends on the Doppler bandwidth. In the bigrating structures treated in this work, we estimate that the figure-of-merit enhancement decreases linearly with the bandwidth (see Appendix~\ref{app:scaling}). Dispersive nanostructures other than the gratings considered here may possess different enhancement-bandwidth scaling. The upper limit of this scaling and thus of the damping enhancement is an open question.

Rather than designing a sail that operates over the full Doppler band, strategies to decrease the operating bandwidth can be employed to activate the resonant effects for the entire acceleration phase. The conceptually simplest method is to tune the laser to higher frequencies during acceleration. This requires $>\SI{200}{\nano\meter}$ bandwidth, which would be extremely difficult given the laser-array scale required for the mission. Another approach is to uniformly stretch the sail, thereby increasing the grating pitch in step with the increasing wavelength. Stretching could be performed using centrifugal forces (\textit{e.g.} sail spinning) on a sail with small Young's modulus or by appropriating the sail heating to promote thermal expansion. Yet, stretching likely cannot accommodate the full 22\% wavelength shift because the maximum elongation is limited by the tensile strength of materials.

Although we concentrate on asymptotic stability, there are other lightsail requirements such as propulsion and thermal management that must be met simultaneously in the membrane design. Doing so requires careful study of the interplay between stability and other metrics. For example, stability through passive optical scattering relies entirely on non-specular orders, whereas propulsion is maximized by specular reflection.
Therefore, it is challenging to simultaneously optimize stability and acceleration in a single architecture~\cite{Salary:2020aa}. 
This is particularly true of purely-reflecting gratings with only first-order diffraction, which we showed cannot simultaneously experience restoring forces and restoring torques.
Regardless, by devising an appropriate figure of merit, we believe that designs with high reflectivity and an optimal tradeoff between acceleration and asymptotic stability can be found~\cite{Salary:2021aa}.

In this work, we chose the incident laser to be TE polarized. However, we are confident that similar damping enhancements can be obtained with gratings optimized for operation with other polarizations. Such gratings will have structures different to those presented here.
The ideal laser polarization in the Starshot project is not yet established, but a promising proposal is that the laser be linearly polarized and rotate synchronously with a spinning sail because spinning greatly benefits beam-riding and flexural stability~\cite{Gao:2024aa}. With this setup, we envision a sail composed of multiple sub-gratings optimized for TE and TM polarizations separately, retaining the advantages of our proposed damping method~\cite{Ilic:2019aa,Gao:2024aa}.

We stress that the dynamic theory developed in Sec.~\ref{sec:Dynamics} and the Appendices can readily be applied to other sail geometries and extended to three dimensions. For different laser-beam profiles such as annular/hollow or multi-modal beams, the analytical framework in this manuscript can be straightforwardly repeated. However, an investigation into the necessity of rigorous relativistic beam transformations~\cite{Yessenov:2023aa} might be required. We anticipate that the effect of the beam transformation is negligible at the relatively low longitudinal and transverse speeds associated with the lightsail mission. 

In our dynamics simulations, we have assumed initial displacements and velocities for the sail, but no further perturbations during its flight. It is likely that the sail is subject to ongoing perturbations, (\textit{e.g.} from imperfectly compensated atmospheric turbulence, laser-beam noise~\cite{Ilic:2019aa,Salary:2020aa} or imperfect laser tracking). During the acceleration phase, these irregularities would generate local distortions in the radiation pressure, which disrupt the local stabilization forces and may accumulate into ejecting the sail from the laser beam. Even in the presence of such continual perturbations, the broadband sail optimized in Sec.~\ref{sec:broadband_grating} could be substantially damped over the acceleration period, especially in combination with the adiabatic reductions in angular coordinates. 
The nature of perturbations must be modeled in order to gauge the damping effectiveness. Moreover, the expected perturbations will inform future optimizations on the best spectral locations for the dispersion enhancement within the Doppler band, such that the largest perturbations are the most damped. For example, one could envisage having the dispersion enhancement at the tail-end of the Doppler band, thus providing the greatest reductions in perturbations at the crucial point where the sail must be accurately pointed at its target. Yet, without damping throughout the sail irradiation period, this strategy risks the sail leaving the laser beam before it reaches the damping-enhancement phase.

The linear stability analysis conducted in Sec.~\ref{sec:Linear Stability Analysis} is adequate for gauging the dynamical behavior of lightsails close to equilibrium (\textit{i.e.} with displacement, angular displacement, velocity and angular velocity near zero). In this regime, the dynamics scales linearly with the initial condition. However, linear stability analysis cannot determine the sail dynamics in regions far from equilibrium. That is, there is a region of state space around equilibrium within which the sail is stable, but sufficient perturbations out of this stable subspace would usher the nonlinear regime. In addition, linear stability analysis does not predict the volume of the stable subspace for a particular laser-and-grating configuration. 
For lightsail missions, the required phase-space volume of the asymptotic-stability region depends on the expected perturbation magnitudes and the properties of the sail. For example, our grating optimization relied on dispersion enhancement near the first-order diffraction cutoff in wavelength. As the wavelength approaches cutoff, the range of sail angles where non-specular grating orders exist tends to zero (as per the grating equation). Therefore, angular perturbations can render the non-specular orders evanescent and destroy asymptotic stability. 
Future optimizations must account for a sufficiently large region around equilibrium for which the sail is asymptotically stable.

One significant assumption applied here and in most of the literature on lightsails~\cite{Lin:2025aa} is that the sail is a rigid body. This assumption is clearly unrealistic because the sail is tens of nanometers thin, the laser beam is imparting nonuniform stresses, and substantial lateral forces are being generated by restoring/damping mechanisms or sail spinning~\cite{Gao:2024aa}. For flexural perturbations with low amplitudes and wavelengths much longer than the nanostructure periodicity, the optical damping suggested in this work is likely to remain because it relies on the scattering properties of bulk periodic structures. However, flexible modes can grow in amplitude from thermal effects or by coupling to radiation pressure~\cite{Dowell:2011aa,Savu:2022aa,Gao:2024aa}, distorting the delicately-tuned membrane and ruining beam-riding stabilization. Methods to minimize flexural perturbations, such as spinning the sail at high-frequencies~\cite{Gao:2024aa}, can help maintain the sail's planar profile and thus protect the optical damping. Even so, radiation-pressure-elastic coupling and relativistic, flexible-membrane beam-riding must be investigated to determine the viability of various stabilizing schemes~\cite{Lin:2025aa}.

Thus, with judicious nanophotonic design, laser-driven sails can experience the full suite of restoring and damping mechanisms in the relativistic regime. The effects are completely integrated into the membrane design at the fabrication stage and are engaged without external input. The design has no active elements, no added internal degrees of freedom and requires no external feedback, simplifying the lightsail design criteria and hence elevating the viability of lightsail missions.

\begin{acknowledgments}
J.Y.L. acknowledges support from an Australian Government Research Training Program (RTP) Scholarship. This research was undertaken with the assistance of resources and services from the National Computational Infrastructure (NCI), which is supported by the Australian Government.
\end{acknowledgments}

\section{Data Availability}
The supporting data and codes for this article are openly available on GitHub~\cite{Lin:2026git}.

\appendix

\section{Special Relativity\label{app:Lorentz Transformation}}
The Lorentz transformation matrix for a simple boost with velocity $\bf v$ is~\cite{Frahm:1979aa}
\begin{align}
    \Lambda(\vb{v})&=
    \begin{pmatrix}
        \gamma              &       - \gamma \beta_x      &       - \gamma \beta_y        &       -\gamma \beta_z         \\
        - \gamma \beta_x    & 1 + \tfrac{\gamma^2}{\gamma+1} \beta_x^2   &  \tfrac{\gamma^2}{\gamma+1} \beta_x \beta_y    &   \tfrac{\gamma^2}{\gamma+1} \beta_x \beta_z  \\
        - \gamma \beta_y    & \tfrac{\gamma^2}{\gamma+1} \beta_x\beta_y   & 1+ \tfrac{\gamma^2}{\gamma+1} \beta_y^2    &   \dfrac{\gamma^2}{\gamma+1} \beta_y \beta_z \\
        - \gamma \beta_z    & \tfrac{\gamma^2}{\gamma+1} \beta_x\beta_z   & \tfrac{\gamma^2}{\gamma+1} \beta_y\beta_z    &   1+\tfrac{\gamma^2}{\gamma+1} \beta_z^2 
    \end{pmatrix}
    \,,
\end{align}
where $\beta_i=v_i/c$, $\gamma \equiv \gamma(\mathbf{v)}$ and we assume $v_z=0$ for a 2D treatment. A four-vector $\vec{A} \xrightarrow[\frameL]{} A^\nu = (A^0, \vb{A})$ has components in frame $\frameM$ defined by a Lorentz boost $\Lambda(\vb{v})$ given by $A^{\mu'}=\Lambda(\vb{v})^{\mu'}_{\nu} A^{\nu}$, or equivalently~\cite{Klacka:2014aa}
\begin{align}
    A^{0'}
    &=
    \gamma \left( A^0 - \bm{\beta} \cdot \vb{A} \right)
    \,, 
    \\
    \vb{A}' 
    &=
    \vb{A} + \bm{\beta} \left[ \frac{\gamma^2}{\gamma+1} \left( \bm{\beta} \cdot \vb{A} \right) - \gamma A^0 \right]
    \,.
\end{align}
For photon emission in direction $\vu{x}$ within frame $\frameL$, the angle of aberration $\theta'$ in frame $\frameM$ is calculated by a Lorentz transformation of the photon four-momentum, yielding
\begin{subequations}\label{eq:Aberration}
\begin{align}
    \sin\theta'(\vb{v}) 
    &= 
    \frac{1}{D(\vb{v})} \left[ -\gamma \beta_y + \frac{\gamma^2}{\gamma+1} \beta_x \beta_y \right] 
    \,,
    \\
    \cos\theta'(\vb{v}) 
    &= 
    \frac{1}{D(\vb{v})} \left[ -\gamma \beta_x + 1 + \frac{\gamma^2}{\gamma+1} \beta_x^2 \right]
    \,.
\end{align}
\end{subequations}

\section{First-order equations of motion\label{app:Derivation}}
We begin from Eq.~\eqref{eq:exact_sail_forces} and obtain the equations of motion linearized to first order in transverse velocity $v_y/c$, angular velocity $\dot{\phi}'L'/c$, frequency offset $\bar{\nu}'$ and angle offset $\bar{\delta}'$, which are all small quantities.

We require the relativistic-velocity addition formula~\cite{RelativisticAddition}
\begin{align}
    \vb{v} \oplus \vb{u}' 
    &= 
    \frac{1}{1+(\vb{v}\cdot\vb{u}')/c^2} \left[ \vb{v} + \frac{\vb{u}'}{\gamma} + \frac{\gamma}{\gamma+1} \frac{(\vb{v}\cdot\vb{u}')}{c^2} \vb{v} \right]
    \,,
\end{align}
where we name $\gamma \equiv \gamma(\mathbf{v)}$ and $D \equiv D(\mathbf{v})$ for convenience. Assuming ${\beta_x\gg\beta_y}$ and non-relativistic lightsail rotation, the velocity addition and subsequent terms reduce to linear order as~\cite{Mackintosh:2024aa}
\begin{subequations}
\begin{align}
    \vb{v} \oplus \vb{u}'&\approx \vb{v} + \frac{u_x'}{\gamma^2} \vu{x} + \frac{u_y'}{\gamma} \vu{y} 
    \,, 
    \\
    \gamma(\vb{v} \oplus \vb{u}') 
    &= 
    \gamma \gamma(\vb{u}') 
    \left[ 1 + \frac{(\vb{v} \cdot \vb{u}')}{c^2} \right] 
    \,, 
    \\
    D(\vb{v} \oplus \vb{u}')&\approx D(\vb{v}) \left( 1 - \frac{u_x'}{c} \right) 
    \,.
\end{align}
\end{subequations}
The transformation from $\frameM(\gyM)$ to $\frameL$ is given by $\Lambda(-\vb{v} \oplus \vb{u}') R(\varepsilon')$, where $R$ is the (2D-embedded) anticlockwise rotation matrix and $\varepsilon'$ is the Thomas-Wigner rotation angle governed by~\cite{WignerAngle}
\begin{subequations} \label{eq:wigner_exact}
\begin{align} 
    \sin\varepsilon'
    &= 
    \frac{(\vb{v} \times \vb{u}')_z}{c^2} 
    \frac{ \gamma \gamma(\vb{u}')(1 + \gamma(\vb{v} \oplus \vb{u}') + \gamma + \gamma(\vb{u}'))}
    {(1+\gamma(\vb{v} \oplus \vb{u}'))(1+\gamma)(1+\gamma(\vb{u}'))} \,, 
    \\
    \cos\varepsilon'
    &=
    \frac{(1 + \gamma(\vb{v} \oplus \vb{u}') + \gamma + \gamma(\vb{u}'))^2}
    {(1+\gamma(\vb{v} \oplus \vb{u}'))(1+\gamma)(1+\gamma(\vb{u}'))} - 1 
    \,.    
\end{align}
\end{subequations}

From Eq.~\eqref{eq:Aberration} and~\eqref{eq:wigner_exact}, we find the linearization of $\theta'(\gyM)$ and $\varepsilon'(\gyM)$ to be 
\begin{subequations} \label{eq:linear_phidot_expansion}
\begin{align}
    \sin\theta'(\gyM) &\approx \sin\theta' - \gyM(\dot{\phi}'/c)A \,, \\
    \cos\theta'(\gyM) &\approx \cos\theta' - \gyM(\dot{\phi}'/c)B \,, \\
    \sin\varepsilon' &\approx - \gyM(\dot{\phi}'/c)\mathcal{E} \,,
\end{align}
\end{subequations}
where, to lowest order in all small quantities,
\begin{subequations} \label{eq:linear_phidot_coefficients}
\begin{align} 
    \Xi &\equiv \beta_x \cos\phi' +\gamma \beta_y \sin\phi' \,,  \\
    \Pi &\equiv \beta_y \cos\phi' +\gamma \beta_x \sin\phi' \,,  \\
    \begin{split}
        A&\equiv \sin\theta' \frac{\cos\phi'}{\gamma D} - \frac{\sin\phi'}{D} + \frac{\Pi}{D(\gamma+1)} \\
        &\hspace{1cm}- \Xi\left[\sin\theta' + \frac{\gamma}{D}\beta_y - \frac{\gamma^2(\gamma+2)}{D(\gamma+1)^2}\beta_x\beta_y \right]  
        \,,
    \end{split} 
    \\
    \begin{split}
        B&\equiv \frac{2 }{D(\gamma+1)} \beta_x \cos\phi' \\
        &\hspace{1cm}- \Xi\left[\cos\theta' + \frac{\gamma}{D}\beta_x - \frac{\gamma^2(\gamma+2)}{D(\gamma+1)^2}\beta_x^2 \right]  
        \,,
    \end{split} 
    \\
    \mathcal{E} &\equiv \frac{\gamma}{\gamma + 1} \left(\sin\phi' {\beta_x} - \cos\phi' \beta_y  \right) 
    \,.
\end{align}
\end{subequations}
The terms $\Xi$ and $\Pi$ are defined for convenience. The factors $A$, $B$ and $\mathcal{E}$ can be interpreted as the linear corrections to the relativistic-aberration sine, relativistic-aberration cosine and the Wigner-angle sine, respectively. When the sail has a non-zero angular velocity, these terms provide the position dependence of the relativistic aberration and Wigner rotation.
From Eq.~\eqref{eq:linear_phidot_coefficients}, we observe that the Thomas-Wigner rotation angle is orders of magnitude smaller than the relativistic aberration angle. Indeed, $\sin\varepsilon'$ has a small factor $\sin\phi'\beta_x$ (from $\mathcal{E}$) multiplying $\gyM\dot{\phi}'/c$, whereas $\sin\theta'$ (Eq.~\eqref{eq:Aberration}) only depends on the small term $\beta_y$. By the same analysis, the aberration correction due to the nonzero angular velocity (the term $\gyM (\dot{\phi}'/c)A$) is orders of magnitude smaller than the base aberration term $\sin\theta'$. Therefore, the angular velocity hardly influences the reference frame orientations and relativistic effects within those frames.

By linearizing the incident angle and frequency on the sail surface in terms of $\gyM$, the angle- and frequency-offset variables can be derived from Eq.~\eqref{eq:offsets} as follows:
\begin{align}
    \bar{\delta}' &\approx \vphi \left[ \frac{A}{\cos\theta'} - \mathcal{E} \right] \,, \\
    \bar{\nu}' &\approx -\vphi D\cos\phi' \,.
\end{align}
Thus, the efficiency factors $\QprM{j}{\prime}(\bar{\delta'},\bar{\nu}')$ can be linearized about $(\bar{\delta'},\bar{\nu}')=\vb{0}$, yielding
\begin{equation} \label{eq:qpr_linearisation}
    \QprM{j}{\prime}(\bar{\delta'},\bar{\nu}') 
    \approx 
    \QprM{j}{\prime}(\delta',\lambda') -\gyM(\dot{\phi'}/c)\TprM{j}{\prime}(\delta',\lambda') 
    \,,
\end{equation}
where
\begin{align}
    \TprM{j}{\prime}
    &\equiv 
    - \left[ \frac{ A }{ \cos\theta' } - \mathcal{E} \right] \frac{\p \QprM{j}{\prime}}{\p \bar{\delta}'} \Bigg\rvert_{\vb{0}} + D \cos\phi' \frac{\p \QprM{j}{\prime}}{\p \bar{\nu}'}\Bigg\rvert_{\vb{0}} 
    \\
    &= 
    \left[ \frac{ A }{ \cos\theta'} - \mathcal{E} \right] \frac{\p \QprM{j}{\prime}}{\p \delta'} + \lambda'  \cos\phi' \frac{\p \QprM{j}{\prime}}{\p \lambda'}
    \,.
\end{align}
The term $\TprM{j}{\prime}$ contains the dispersive contributions of the radiation-pressure efficiency factors. The second line follows from the first by reformulating the derivatives with respect to $\bar{\delta}'$ and $\bar{\nu}'$ into derivatives with respect to the angle or wavelength at the CoM, which are more convenient variables in electromagnetic simulations.

Inserting Eq.~\eqref{eq:qpr_linearisation} into Eq.~\eqref{eq:exact_sail_forces}, we find the total force and torque on the sail linearized in terms of angular velocity:
\begin{widetext}
\begin{subequations}\label{eq:linear_forces}
\begin{align}
    F_{x}' 
    &\approx 
    \int_{-L'/2}^{L'/2} dg_2' \, 
    D^2(\vb{v}) \frac{I(g_2')}{c} 
    \Bigg\{
        \left[ \cos\theta' \QprM{1}{\prime} - \sin\theta' \QprM{2}{\prime} \right] \notag \\
        &\hspace{50pt}+ \vphi 
        \bigg( 
            \left[ 2\cos\phi' \QprM{1}{\prime} - \TprM{1}{\prime} \right] \cos\theta' - \left[ 2\cos\phi' \QprM{2}{\prime} - \TprM{2}{\prime} \right] \sin\theta' \notag \\
            &\hspace{110pt}
            - \left(B - \mathcal{E}\sin\theta' \right) \QprM{1}{\prime} + \left(A + \mathcal{E}\cos\theta' \right) \QprM{2}{\prime}   
        \bigg)
    \Bigg\}(\delta',\lambda') 
    \,, \\
    F_{y}'
    &\approx 
    \int_{-L'/2}^{L'/2} dg_2' \, 
    D^2(\vb{v}) \frac{I(g_2')}{c} 
    \Bigg\{
        \left[ \sin\theta' \QprM{1}{\prime} + \cos\theta' \QprM{2}{\prime} \right] \notag \\
        &\hspace{50pt}+ \vphi 
        \bigg( 
            \left[ 2\cos\phi' \QprM{1}{\prime} - \TprM{1}{\prime} \right] \sin\theta' + \left[ 2\cos\phi' \QprM{2}{\prime} - \TprM{2}{\prime} \right] \cos\theta' \notag \\
            &\hspace{110pt}
             -  \left(A + \mathcal{E}\cos\theta' \right) \QprM{1}{\prime} - \left(B - \mathcal{E}\sin\theta' \right) \QprM{2}{\prime}  
         \bigg)
    \Bigg\}(\delta',\lambda') 
    \,, \\
    \boldsymbol{\tau}'
    &\approx 
    - \vu{g}_3' 
    \int_{-L'/2}^{L'/2} dg_2' \, 
    \gyM D^2(\vb{v}) \frac{I(g_2')}{c}
    \Bigg\{
        \left[ \cos\delta' \QprM{1}{\prime} - \sin\delta' \QprM{2}{\prime} \right] \notag \\
        &\hspace{50pt}+ \vphi 
        \bigg( 
            \left[ 2\cos\phi' \QprM{1}{\prime} - \TprM{1}{\prime} \right] \cos\delta' - \left[ 2\cos\phi' \QprM{2}{\prime} - \TprM{2}{\prime} \right] \sin\delta' \notag \\
            &\hspace{110pt}
            - \left(C - \mathcal{E}\sin\delta' \right) \QprM{1}{\prime} + \left(S + \mathcal{E}\cos\delta' \right) \QprM{2}{\prime}   
        \bigg)
    \Bigg\}(\delta',\lambda') \,,
\end{align}
\end{subequations}
\end{widetext}
where $S \equiv A\cos\phi' - B\sin\phi'$ and $C \equiv A\sin\phi' + B\cos\phi'$. These equations apply in frame $\frameM$, no longer involving frames $\frameM(\gyM)$. In particular, notice that all terms in curly braces depend on $\delta'$ and $\lambda'$, which are the angle and wavelength measured at the CoM in $\frameM$.

%%%
\section{Linear stability\label{app:Linear Stability}}
%%%
In this section, we derive the equations of motion upon linearizing Eq.~\eqref{eq:linear_forces} in terms of the transverse variables $\mathbf{q}'' = [y'', \phi'', v_y'', \dot{\phi}'']$. Then, we analyze the analytic eigenvalues of the linear-stability Jacobian Eq.~\eqref{eq:JacobianMatrix} in the small-perturbation regime.

%In the linear regime, we approximate the form of the Gaussian laser intensity. 
We assume that transverse coordinates between frames $\frameU$ and $\frameL$ are equal, \textit{i.e.} $y''=y$. This ceases to be an approximation when $v_y=0$ and ensures the Gaussian-beam intensity profile is explicitly time independent when $v_y\neq 0$. The resultant Gaussian laser-beam intensity is 
\begin{align}
    I(\gyU) &= I_0 \exp\left[-2 \left( \frac{y'' + \gyU \cos\phi''}{w} \right)^2 \right]. 
\end{align}

The left-right mirror symmetry of the sail when $\mathbf{q}''=\mathbf{0}$ results in the radiation-pressure efficiency symmetries
\begin{align}
    \left[\QprM{2}{R\prime\prime} + \QprM{2}{L\prime\prime} \right](\mathbf{q}''=\vb{0}) 
    &= 
    0 \,, \label{eq:qpr2_balance}
    \\ 
    \left[\QprM{1}{R\prime\prime} - \QprM{1}{L\prime\prime} \right](\mathbf{q}''=\vb{0}) 
    &= 0 \,.
    \label{eq:qpr1_balance}
\end{align}
Equation~\eqref{eq:qpr2_balance} encodes the transverse forces on the two sail halves balancing at $\mathbf{q}''=\mathbf{0}$, ensuring the sail is not accelerating in the $y''$-direction at $\mathbf{q}''=\mathbf{0}$. Equation~\eqref{eq:qpr1_balance} shows that the longitudinal forces on each half are equal at $\mathbf{q}''=\mathbf{0}$, which prevents the sail from experiencing a torque at $\mathbf{q}''=\mathbf{0}$. There are similar symmetry relations for the derivatives of the efficiency factors with respect to angle $\delta'$ and wavelength $\lambda'$ at equilibrium. Therefore, $\mathbf{q}''=\mathbf{0}$ is an equilibrium of the transverse subspace.

Upon linearizing Eq.~\eqref{eq:linear_forces} to first order in $\mathbf{q}''$ at $\mathbf{q}''=\mathbf{0}$, we obtain the linear equations of motion
\begin{subequations} \label{eq:linear_eom}
\begin{align}
    F_y''/m &= \kyy y'' + \kyp \phi'' + \myp \dot{\phi}'' + \myy v_y'' \,,
    \\
    \tau''/J'' &= \kpy y'' + \kpp \phi'' + \mpp \dot{\phi}'' + \mpy v_y'' \,.
\end{align}
\end{subequations}
The coefficients in these equations are determined by the linearization and comprise the Jacobian matrix Eq.~\eqref{eq:JacobianMatrix}. Given the symmetries of Eqs.~\eqref{eq:qpr2_balance} and~\eqref{eq:qpr1_balance}, we may express the Jacobian coefficients in terms of the right-half efficiency factors $\QprM{j}{R\prime}$ only. The result is
\begin{subequations}\label{eq:Jacobian Terms}
\begin{align}
    \kyy(\lambda') 
    &=
    - D^2 \frac{2P_0}{mc} \frac{1}{L'} \winf \QprM{2}{R\prime}(\lambda') 
    \,, \\
    \kpy(\lambda') 
    &=
    D^2 \frac{2P_0}{mc} \frac{1}{\Gamma L'^2} \wf{1/2} \QprM{1}{R\prime} (\lambda')
    \,, \\
    \kyp(\lambda') 
    &=
    - D^2 \frac{2P_0}{mc} \ws{1/2} \frac{\p \QprM{2}{R\prime}}{\p \delta'} (\lambda')
    \,, \\
    \kpp(\lambda') 
    &=
    - D^2 \frac{2P_0}{mc} \frac{1}{\Gamma L'} \ws{0} 
    \left[ 
        \QprM{2}{R\prime} - \frac{\p \QprM{1}{R\prime}}{\p \delta'}
    \right]\! (\lambda')    
    \,, \\
    \myy(\lambda')
    &=
    - D^2 \frac{2P_0}{mc} \frac{1}{c} \frac{D+1}{D(\gamma+1)} \ws{1/2} 
    \left[ 
        \QprM{1}{R\prime} + \frac{\p \QprM{2}{R\prime}}{\p \delta'}
    \right]\! (\lambda') 
    \,, \\
    \mpy(\lambda') 
    &= 
    \frac{1}{c} \frac{D+1}{D(\gamma+1)} \kpp(\lambda') 
    \,, \\
    \myp(\lambda') 
    &=
    D^2 \frac{2P_0L'}{mc} \frac{1}{c} \ws{0} 
    \left[
        2 \QprM{2}{R\prime} - \lambda' \frac{\p  \QprM{2}{R\prime}}{\p \lambda'}
    \right]\! (\lambda')
    \,, \\
    \mpp(\lambda')
    &=
    - D^2 \frac{2P_0}{mc} \frac{1}{c} \frac{1}{\Gamma} \wf{0} 
    \left[
        2 \QprM{1}{R\prime} - \lambda' \frac{\p  \QprM{1}{R\prime}}{\p \lambda'}
    \right]\! (\lambda') 
    \,.
\end{align}
\end{subequations}
These expressions contain coefficients that depend only on the Gaussian-beam width, which are defined as
\begin{subequations} \label{eq:Jacobian_width}
\begin{align}
    \winf &\equiv \sqrt{\frac{2}{\pi}} \frac{1}{\bar{w}} \bigg[ 1 -  \exp\bigg(-\frac{1}{2\bar{w}^2} \bigg) \bigg] \,, \\
    \ws{1/2} &\equiv \frac{1}{2} \erf \bigg(\frac{1}{\bar{w} \sqrt{2}}\bigg) \,, \\
    \wf{1/2} &\equiv \ws{1/2} - \sqrt{\frac{2}{\pi}} \frac{1}{2\bar{w}}  \exp\bigg(-\frac{1}{2\bar{w}^2} \bigg) \,, \\
    \ws{0} &\equiv \frac{\bar{w}^2}{4} \winf \,, \\
    \wf{0} &\equiv \frac{\bar{w}^2}{4} \wf{1/2} \,,
\end{align}
\end{subequations}
where $\bar{w} \equiv w/L'$. All of these are positive and tend to zero as $\bar{w} \rightarrow \infty$; the superscript (if any) indicates how quickly the factors approach zero in this limit, whereas the subscript indicates the limit value as $\bar{w} \rightarrow 0$.

\subsection{First-order perturbation theory\label{app:perturbation}}
Since the damping terms are much smaller than the restoring terms (by roughly a factor $c$), we treat the Jacobian as a sum of an unperturbed and perturbative Jacobian, respectively:
\begin{align}
    \vb{J}_\text{unp}
    \equiv
    \begin{pmatrix}
        0 & 0 & 1 & 0 \\
        0 & 0 & 0 & 1 \\
        \kyy & \kyp & 0 & 0 \\
        \kpy & \kpp & 0 & 0
    \end{pmatrix}
    \,, \quad
    \delta\vb{J}
    \equiv
    \begin{pmatrix}
        0 & 0 & 0 & 0 \\
        0 & 0 & 0 & 0 \\
        0 & 0 & \myy & \myp \\
        0 & 0 & \mpy & \mpp
    \end{pmatrix}
    \,.
\end{align}
The eigenvalues of the unperturbed matrix are calculated straightforwardly:
\begin{subequations} \label{eq:unpert_eigval}
\begin{align}
    \xi_{\text{unp},1,2}
    &=
    \pm \frac{1}{\sqrt{2}} 
    \Big[ 
        \kyy + \kpp - \sqrt{(\kyy - \kpp)^2 + 4\kpy\kyp}
    \Big]^{1/2}
    \,,
    \\
    \xi_{\text{unp},3,4}
    &=
    \pm \frac{1}{\sqrt{2}} 
    \Big[ 
        \kyy + \kpp + \sqrt{(\kyy - \kpp)^2 + 4\kpy\kyp}
    \Big]^{1/2}
    \,.
\end{align}
\end{subequations}
These unperturbed eigenvalues can have nonzero real part, as expected of the eigenvalues of real matrices. Due to the $\pm$ signs, the real parts come in $\pm$ pairs, \textit{i.e.} any damped mode is accompanied by an exponentially growing mode. Therefore, explicit damping terms $\mu$ are necessary for all modes to decay simultaneously.

The aim is to solve the full eigenvalue problem $\mathbf{J} \mathbf{v} = \xi \mathbf{v}$, where $\mathbf{v} = \mathbf{v}_\text{unp} + \delta\mathbf{v}$ and $\xi = \xi_\text{unp} + \delta\xi$ are the true eigenvectors and eigenvalues of $\mathbf{J}$, respectively. Assuming all eigenvalues have multiplicity 1, we expand the eigenvalue equation and discard $\delta$ quantities with second or higher-order, yielding
\begin{equation} \label{eq:pert_eigval}
    \delta\xi_i = \frac{\mathbf{w}_i (\delta\mathbf{J}) \mathbf{v}_i}{\mathbf{w}_i \mathbf{v}_i} \,,
\end{equation}
where $\mathbf{v}_i$ is the $i$-th right-eigenvector of $\mathbf{J}_\text{unp}$ and $\mathbf{w}_i$ is the $i$-th left-eigenvector ($\mathbf{w}_i \mathbf{J}_\text{unp} = \xi_\text{unp} \mathbf{w}_i$). Equation~\eqref{eq:pert_eigval_jac} for the perturbative eigenvalues of $\mathbf{J}$ can thus be derived from Eq.~\eqref{eq:pert_eigval}.

The eigenvalues of Eq.~\eqref{eq:pert_eigval_jac} can be interpreted more easily in a special case. First, assume $\kyy \kpp > \kyp \kpy$, which is a necessary condition for the Jacobian to have strictly negative real-part eigenvalues according to the Routh-Hurwitz stability criteria~\cite{RouthTextbook}. If we further assume that $\kyy + \kpp < 0$ (the sail has at least a strong restoring force or torque), then it is easy to see that $\xi_{\text{unp},1\text{--}4}$ are purely imaginary and that $\delta\xi_{1\text{--}4}$ are purely real.
That is, under these somewhat common conditions, the unperturbed eigenvalues form the imaginary/oscillatory part of $\xi_{1\text{--}4}$ and the perturbation leads to damping.

Further simplification occurs when $(k^y_y-k^\phi_\phi)^2 \gg |4k^y_\phi k^\phi_y|$, which is frequently satisfied in gratings found through asymptotic stability optimization (using Eq.~\eqref{eq:FOM}). In this case, we can take a first-order Taylor expansion of the denominator in Eq.~\eqref{eq:pert_eigval} to simplify the eigenvalues to the following:
\begin{subequations} \label{eq:pert_eigval_decoupled}
\begin{align} 
    \Re(\xi_\text{dominant}) 
    &\approx 
    \frac{1}{2} \left[\mu^y_y - \frac{k^y_\phi \mu^\phi_y + k_y^\phi \mu_\phi^y}{k^\phi_\phi - k_y^y} \right]
    \,,
    \\
    \Re(\xi_\text{weak}) 
    &\approx 
    \frac{1}{2} \left[\mu^\phi_\phi + \frac{k^y_\phi \mu^\phi_y + k_y^\phi \mu_\phi^y}{k^\phi_\phi - k_y^y} \right]
    \,,
\end{align}
\end{subequations}
which decouples the translational and rotational damping coefficients. These equations associate the weak eigenvalue with the rotational damping and the dominant eigenvalue with the translational damping, which was observed in Sec.~\ref{sec:narrow_band}. Due to the magnitude disparity between the rotational and translational damping, the cross-coupling term in Eq.~\eqref{eq:pert_eigval_decoupled} strongly influences the dominant eigenmode, but not the weak eigenmode.

\section{Diffraction-grating calculations\label{app:grating}}
From ray-momentum-transfer arguments, the radiation-pressure efficiency factors for a diffraction grating are~\cite{Lin:2024aa}:
\begin{align}
    \QprM{1}{\prime}(\delta',\lambda')
    &= 
    \cos\delta' 
    \bigg\{1 + \sum_m \Big[ r_m' (\delta',\lambda') \cos (\delta'+\delta_m') \notag 
    \\
    &\hspace{50pt}- t_m'(\delta',\lambda') \cos(\delta'-\delta_m') \Big] \bigg\} 
    \,, \label{eq:Qpr1_bigrating} 
    \\
    \QprM{2}{\prime}(\delta',\lambda')
    &= 
    -\cos\delta' 
    \sum_m \Big[ r_m' (\delta',\lambda') \sin(\delta'+\delta_m') \notag 
    \\
    &\hspace{50pt}- t_m'(\delta',\lambda')\sin(\delta'-\delta_m') \Big] 
    \,. \label{eq:Qpr2_bigrating}
\end{align}

For an absorption-free grating, we have the following energy conservation condition~\cite{Maystre:2014aa}
\begin{equation} \label{eq:energy_cons}
    \sum_{m=-\infty}^\infty \big[ r_m'(\delta',\lambda') + t_m(\delta',\lambda') \big] = 1 
    \,,
\end{equation}
and for purely-reflecting gratings ($t_m = 0$ for all $m\in\mathbb{Z}$), the reciprocity theorem produces:
\begin{equation} \label{eq:reciprocity}
    \refl{L}{0}(\delta',\lambda') = \refl{R}{0}(\delta',\lambda') \,.
\end{equation}
This equation is nontrivial because it applies even when the unit cell is asymmetric, which is true of the bigratings considered in this manuscript.

Assuming a bigrating with left-right mirror symmetry, the efficiencies can be expressed in terms of the left or right half using the following conditions:
\begin{subequations} \label{eq:LR_symmetry}
\begin{align} 
    [r,t]^{R\prime}_m(\delta',\bar{\nu}') 
    &= 
    [r,t]^{L\prime}_{-m}(-\delta',\bar{\nu}') 
    \,,
    \\
    \frac{\p}{\p\delta'}[r,t]^{R\prime}_m(\delta',\bar{\nu}') 
    &= 
    -\frac{\p}{\p\delta'}[r,t]^{L\prime}_{-m}(-\delta',\bar{\nu}') 
    \,,
    \\
    \frac{\p}{\p\bar{\nu}'}[r,t]^{R\prime}_m(\delta',\bar{\nu}') 
    &= 
    \frac{\p}{\p\bar{\nu}'}[r,t]^{L\prime}_{-m}(-\delta',\bar{\nu}') 
    \,.
\end{align}
\end{subequations}

\section{Numerical optimization\label{app:Numerical optimisation}}

\subsection{Method}
Maxwell's equations for the diffraction grating were solved using rigorous coupled-wave analysis in the TORCWA Python package~\cite{kim:2023torwa}, with PyTorch~\cite{Paszke:2019aa} used to perform gradient calculations. We took the incident plane wave to have TE polarization and $\lzero = \SI{1}{\micro\meter}$. When optimizing lightsails aiming for $v_f=0.2c$, the average over wavelength in Eq.~\eqref{eq:FOM} is performed over the wavelength range $[\lzero, 1.22\lzero]$ corresponding to the total Doppler shift, which amounts to just over \SI{200}{\nano\meter} of bandwidth. We assumed that spatial intensity variations of the Gaussian beam occur on a much larger length scale than that of the grating period, which is reasonable given that the beam width is of similar size to the macroscopic sail diameter. Hence, the diffraction efficiencies accurately represent the proportion of scattered power on each cluster of unit cells. 

\begin{table}[!htb]
    \centering
    \begin{tabular}{c|c}
         Parameter & Bounds 
         \\ \hline
         % Grating pitch, $\Lambda'$ & ($\lmax/(1 - \sin\SI{0.1}{\degree})$, $\lmax/(1 + \sin\SI{15}{\degree})$) 
         % \\ 
         Grating pitch, $\Lambda'$ & $\lmax/(1 - \sin\SI{0.1}{\degree})$ 
         \\ 
         Grating thickness, $h'$ [$\Lambda'$] & ($0.01$, $1.5$) 
         \\ 
         Resonator widths, $w'$ [$\Lambda'$] & ($0.01$, $1$) 
         \\
         Resonator-center separation, $\Delta x'$ [$\Lambda'$] & ($0.03$, $0.5$)
         \\
         Resonator permittivities, $\epsilon_{1,2}'$ & ($1.1^2$, $3.5^2$) 
         \\
         Gaussian-beam width, $w$ [$L'$] & 2
         \\
         Substrate thickness, $h_\text{sub}'$ [$\Lambda'$] & ($0.01$, $1.5$) 
         \\
         Substrate permittivity, $\epsilon_\text{sub}'$ & ($1.1$, $3.5$)
    \end{tabular}
    \caption{Upper and lower bounds or fixed values for the optimization search.}
    \label{tab:bounds}
\end{table}

We used multistart multi-level single-linkage~\cite{Rinnooy-Kan:1987ab} as a global point sampler coupled with the method of moving asymptotes~\cite{Svanberg:1987aa} to perform local gradient descent, both available in the NLopt Python package~\cite{Johnson:2007aa}. The global point sampler searches the parameter space with bounds defined in Tab.~\ref{tab:bounds}, which we discuss as follows. For all optimizations except that at a single-wavelength [Sec.~\ref{sec:narrow_band}], we fixed the pitch equal to the expression in the first row of Tab.~\ref{tab:bounds} to utilize grating-cutoff effects and reduce the size of the search space. In contrast to our previous approach~\cite{Lin:2024aa}, the grating pitch is optimized instead of the laser wavelength. This choice enables efficient gradient calculation for the figure of merit [Eq.~\eqref{eq:FOM}], which contains an average over wavelength. Across all parameters, the minimum bound is offset from 0 (or 1 for the permittivities) to avoid finding structures with zero Jacobian determinant. The thicknesses are kept reasonably small given the necessarily low sail mass. To largely avoid searching redundant parameter regions, resonator widths and center separation are upper bounded by the grating pitch and half-pitch, respectively. Furthermore, inequality constraints on these parameters were passed to the local optimizer.  However, the global optimization routine does not respect the local optimization constraints and sometimes initializes points in the redundant parameter spaces. The Gaussian-beam width can also be added as an optimization parameter, but was fixed in our final optimizations. After several runs, we found $w = 2L'$ was a reasonable tradeoff between an infinitely narrow beam ($w \rightarrow 0$) and a plane wave ($w \rightarrow \infty$). Finally, the permittivities are varied between vacuum and silicon, which is the typical range for proposed lightsail materials~\cite{Lin:2025aa}.

\begin{table}[!htb]
    \centering
    \begin{tabular}{c|c|c|c|c}
         Parameter & NB & NB sens. [\%] & BB & BB sens. [\%]
         \\ \hline
         $\Lambda'$ [\SI{}{\micro\meter}] & $1.328$ & $\SI{2.5e3}{}$ & $1.227$ & $-1.9$
         \\ 
         $h'$ [$\Lambda'$] & $1.649$ & $\SI{4.6e1}{}$ & $0.2308$ & $0.18$
         \\ 
         $w_1'$ [$\Lambda'$] & $1.101$ & $\SI{4.5e2}{}$ & $0.3196$ & $-0.014$
         \\ 
         $w_2'$ [$\Lambda'$] & $1.169$ & $\SI{3.3e2}{}$ & $0.6325$ & $-0.071$
         \\
         $\Delta x'$ [$\Lambda'$] & $ 0.2631$ & $-\SI{1.5e2}{}$ & $0.1728$ & $0.0011$
         \\
         $\epsilon_{1}'$ & $7.435$ & $\SI{9.6e2}{}$ & $6.060$ & $-0.014$
         \\
         $\epsilon_{2}'$ & $4.020$ & $\SI{6.4e2}{}$ & $4.209$ & $-0.14$
         \\
         $w$ [m] & $20$ & $0.083$ & $20$ & $-0.062$
         \\
         $h_\text{sub}'$ [$\Lambda'$] & $0.7561$ & $\SI{1.1e1}{}$ & $0.2708$ & $0.023$
         \\
         $\epsilon_\text{sub}'$ & $6.618$ & $\SI{1.4e1}{}$ & $2.895$ & $-0.065$
    \end{tabular}
    \caption{Asymptotic-stability-optimized diffraction grating unit-cell parameters. The narrow-band grating (NB) and broadband grating (BB) are discussed in Sections~\ref{sec:narrow_band} and~\ref{sec:broadband_grating}, respectively. The associated sensitivity (sens.) to a 0.1\% parameter variation is also shown.}
    \label{tab:gratings}
\end{table}

The final parameter values for the gratings discussed in the main text are shown in Tab.~\ref{tab:gratings} columns~2 and~4. We remark that the resonator widths ($w_1'$, $w_2'$) and permittivities ($\epsilon_{1}'$, $\epsilon_{2}'$) shown in this table do not necessarily reflect the physical widths or permittivities of the resonators because the resonators are allowed to overlap. That is, the distance between resonator centers, $\Delta x'$, may be smaller than $(w_1' + w_2')/2$ (see Fig.~\ref{fig:structure}(a)), which is the case for both the narrow-band and broadband gratings in Tab.~\ref{tab:gratings}. Instead, the tabulated values are inputs in our grating implementation~\cite{Lin:2026git}; the actual grating and permittivity profiles are shown in Fig.~\ref{fig:structure}(b). Allowing the parameters to vary smoothly in this manner ensures gradients are calculated correctly during optimization.

\subsection{Sensitivity analysis}
We gauge the sensitivity of the figure of merit $F_\text{asymp}$ to variations in the parameters $p$. One estimate comes from the gradients $\partial F_\text{asymp}/\partial p$, which are obtained naturally during the optimization. We evaluate the relative change in $F_\text{asymp}$ from the optimum value when varying the parameters by a given percentage $\Delta p/p$: $\Delta F_\text{asymp}/F_\text{asymp} = (p/F_\text{asymp})(\partial F_\text{asymp}/\partial p) (\Delta p/p)$. In choosing $\Delta p/p$, a value of 0.1\% is reasonable given the fabrication precision of gratings with around \SI{1}{\micro\meter} pitch. Therefore, for $\Delta p/p = 0.1\%$ variation in each of the parameters, the corresponding variations in $F_\text{asymp}$ are shown in Tab.~\ref{tab:gratings} columns~3 and~5. The variation in the weak eigenvalue (\textit{i.e.} the eigenvalue not captured by $F_\text{asymp}$) has a similar order of magnitude to the dominant eigenvalue and is not shown. 
% The wavelength range is determined by the final speed of the craft, which in turn depends on the mission requirements. Over finite ranges, designs that do not rely on sharp resonances and are thus robust to perturbations can be found.

\begin{figure}
    \centering
    \includegraphics[width=0.9\linewidth]{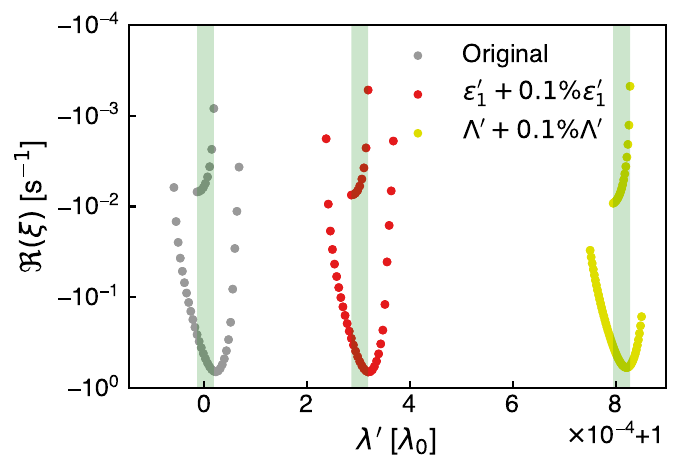}
    \caption{The narrow-band grating [Fig.~\ref{fig:structure}(bi)] resonance shifts away from $\lzero=\SI{1}{\micro\meter}$ with variations in pitch or permittivity, but retains its magnitude. The green shading denotes the wavelength regions where all eigenvalues have negative real part.}
    \label{fig:resonance_shift}
\end{figure}

From Tab.~\ref{tab:gratings} column~5, the broadband grating is clearly robust against deviations in the parameters. 
For the narrow-band grating, the performance is seemingly very sensitive to variations in the grating parameters because the enhancement relies entirely on a sharp resonance. However, the resonance itself is somewhat resilient to changes in geometrical or optical parameters. For example, if the grating pitch or a unit-cell-resonator permittivity is increased by 0.1\%, the resonance retains its magnitude but shifts to a wavelength different from $\lzero = \SI{1}{\micro\meter}$ [Fig.~\ref{fig:resonance_shift}]. Varying the other grating parameters results in similar behavior. Therefore, the narrow-band grating can still function despite minor deviations in geometric or optical properties that arise from fabrication imperfections.

The sensitivities of the broadband and narrow-band gratings were also observed in the optimization results. The narrow-band gratings were often highly isolated local minima, while the broadband gratings resided in relatively robust, stable regions of the optimization parameter space.

\section{Enhancement scaling with bandwidth\label{app:scaling}}
\begin{figure}
    \centering
    \includegraphics[width=0.8\linewidth]{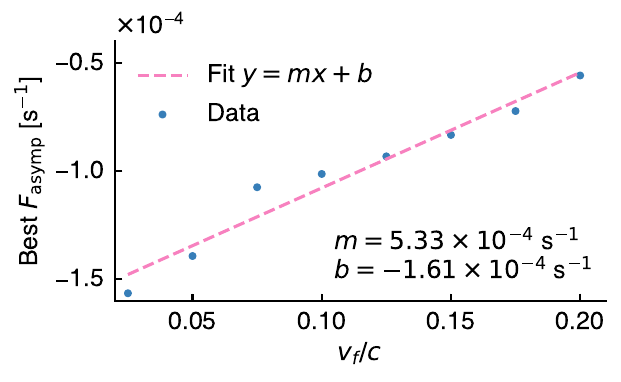}
    \caption{Optimal damping enhancement scaling with final velocity (thus laser bandwidth). Each point corresponds to a (possibly different) optimal grating that was optimized for a specific final velocity.}
    \label{fig:scaling}
\end{figure}

To estimate the tradeoff between Doppler bandwidth and damping enhancement, we repeated the optimization of Eq.~\eqref{eq:FOM} over wavelength ranges corresponding to final velocities from $0.025c$ to $0.2c$ in increments of $0.025c$. Each optimization was conducted under the same conditions as that in Sec.~\ref{sec:broadband_grating}, but with a different final-velocity target. In Fig.~\ref{fig:scaling}, we plot the best $F_\text{asymp}$ (smaller is better) against the final velocity (corresponding to the Doppler bandwidth), showing that the damping magnitude decreases linearly with bandwidth. As the Doppler bandwidth approaches zero (not shown), the scaling is likely nonlinear because the damping relies more on narrow-band resonant enhancement. For instance, if the area under the damping resonance is fixed, then the damping enhancement corresponding to the peak value would scale inversely with the resonance bandwidth.

\section{Dynamics\label{app:Dynamics}}
The equations of motion [Eq.~\eqref{eq:linear_forces}] were integrated using the comoving solver discussed in Sec.~\ref{sec:Dynamics}, stopping when the sail reached $v_f = 0.2c$. Within the comoving integrator, the frame-$\frameM$ equations used to evolve the state were solved using a fourth-order Runge-Kutta method. A detailed implementation of the dynamics solver is stored in the code repository~\cite{Lin:2026git}.

To quantify the range of initial conditions for which the broadband grating of Sec.~\ref{sec:broadband_grating} is stable, we conducted further dynamics simulations. The initial longitudinal position and velocity were fixed to zero in all cases ($[x,v_x] = [0,0]$). Each simulation had only one nonzero initial coordinate, which we repeated for three different values (\textit{e.g.} $[y_0,\phi_0',v_{y,0},\dot{\phi}_0'] = [0.02w,0,0,0], [0.04w,0,0,0]$ and $[0.06w,0,0,0]$). For every initial coordinate, we only probed positive perturbations because the initial condition is symmetric with respect to zero when only one coordinate is perturbed. The divergence criteria for these simulations was whether the angle $\phi'$ exceeded the first-order grating cutoff at $\lambda'=\lzero$ (about $\ang{10.6}$). We estimated the upper limits of the stable initial conditions based on whether $\phi'$ exceeded $\ang{10.6}$ within the first 2~minutes of integration. If so, the equations of motion were integrated until the sail reached $0.2c$. 
The sail remained stable for $y_0 = 0.02w, 0.04w, 0.06w$, $\phi_0' = \ang{2}, \ang{5}, \ang{8}$, $v_{y,0} = \SI{1}{\meter\per\second}, \SI{2}{\meter\per\second}, \SI{3}{\meter\per\second}$ or $\dot{\phi}_0' = \SI{0.3}{\rps}, \SI{0.6}{\rps}, \SI{1}{\rps}$ individually. These ranges are a rough upper bound on initial conditions for which the sail is stable.

\bibliography{references}% Produces the bibliography via BibTeX.
\end{document}